\documentclass[ras]{agutex}

\usepackage[dvips]{graphicx}
\authorrunninghead{HELMBOLDT ET. AL}

\titlerunninghead{VLA-BASED SPECTRAL ANALYSIS OF TEC FLUCTUATIONS}

\authoraddr{J.\ F.\ Helmboldt, US Naval Research Laboratory, 
Code 7213, 4555 Overlook Ave. SW, Washington, DC 20375 
(joe.helmboldt@nrl.navy.mil)}

\authoraddr{T.\ J.\ W.\ Lazio, Jet Propulsion Laboratory, California Institute of Technology, M/S 138-308, 4800 Oak Grove Dr., Pasadena, CA 91106
(joseph.lazio@jpl.nasa.gov)}

\authoraddr{H.\ T.\ Intema, National Radio Astronomy Observatory, 
520 Edgemont Rd, Charlottesville, VA 22903 
(hintema@nrao.edu)}

\authoraddr{K.\ F.\ Dymond, US Naval Research Laboratory, 
Code 7655, 4555 Overlook Ave. SW, Washington, DC 20375 
(kenneth.dymond@nrl.navy.mil)}

\begin{document}

\title{A New Technique for Spectral Analysis of Ionospheric TEC Fluctuations Observed with the Very Large Array VHF System: From QP Echoes to MSTIDs}

\authors{J. F. Helmboldt, \altaffilmark{1} T. J. W. Lazio, \altaffilmark{2}  H. T. Intema, \altaffilmark{3} \& K. F. Dymond, \altaffilmark{1}}

\altaffiltext{1}{US Naval Research Laboratory, Washington, DC, USA.}
\altaffiltext{2}{Jet Propulsion Laboratory, California Institute of Technology, Pasadena, CA, USA.}
\altaffiltext{3}{Jansky Fellow of the National Radio Astronomy Observatory, Charlottesville, VA, USA.}

\begin{abstract}
We have used a relatively long, contiguous VHF observation of a bright cosmic 
radio source (Cygnus A) with the Very Large Array (VLA) through the nighttime, 
midlatitude ionosphere to demonstrate the phenomena observable with this instrument.  In a companion paper, we showed that the VLA can detect fluctuations in total 
electron content (TEC) with amplitudes of $\leq\!10^{-3}$ TECU and can measure TEC gradients with a precision of about $2 \times 10^{-4}$ TECU km$^{-1}$.  
We detail two complementary techniques for producing spectral 
analysis of these TEC gradient measurements.  The first is able to track individual waves 
with wavelengths of about half the size of the array ($\sim 20$ km) or more.  
This technique was successful in detecting and characterizing many medium-scale 
traveling ionospheric disturbances (MSTIDs) seen intermittently throughout the 
night and has been partially validated using concurrent GPS measurements.  
Smaller waves are also seen with this technique at nearly all times, 
many of which move in similar directions as the detected MSTIDs.  The second 
technique allows for the detection and statistical description of the 
properties of groups of waves moving in similar directions with wavelengths 
as small as 5 km.  Combining the results of both spectral techniques, we found 
a class of intermediate and small scale waves which are likely the 
quasi-periodic (QP) echoes that have been observed to occur within 
sporadic-$E$ ($E_s$) layers.  We find two distinct populations of these waves.  
The members of one 
population are coincident in time with MSTIDs and are consistent with 
being generated within $E_s$ layers by the $E$--$F$ coupling instability.  
The other population seems more influenced by the neutral wind, similar to the 
predominant types of QP echoes found by the Sporadic-$E$ Experiments over Kyushu \citep[SEEK][]{fuk98,yam05}.  We have 
also found that the spectra of background (i.e., isotropic) fluctuations 
can be interpreted as the sum of two turbulent components with maximum 
scales of about 300 km and 10 km.
\end{abstract}

\begin{article}

\section{Introduction}
Ground-based remote sensing of the ionosphere is a rich field of study that has developed over several decades and now includes a host of different instruments including, but not limited to, GPS, CERTO \citep{ber06}, and other radio beacon receivers, ionosondes, and radar/HF arrays.  A powerful yet relatively underused resource for remote sensing are radio-frequency arrays, particularly those that operate in the VHF regime.  Designed as radio synthesis telescopes, they are chiefly used to observe cosmic sources.  While such observations require detailed calibration schemes to remove the effects of the ionosphere, this calibration data is seldom used to actually study the ionosphere.\par
These interferometers basically measure the time-averaged correlation of complex voltages from pairs of antennas which can be combined with Fourier methods to make relatively high angular resolution images of cosmic sources \citep[for a detailed discussion of the methods involved, see][]{tho91}.  The correlated signals, or ``visibilities'' have an extra phase term added to them by the ionosphere proportional to the difference in the total electron content (TEC) between the two antennas' lines of sight.  Because of the relatively large collecting area of the individual elements (usually dishes, dozens of meters across), and the brightness of many cosmic sources, these additional phase terms can typically be converted to differential TEC ($\delta \mbox{TEC}$) measurements with a precision of $\sim\!10^{-3}$ TECU or better.  In addition, the range of scales that can be probed with such interferometers (dozens of meters to hundreds of kilometers) and the virtual ubiquity of target sources available on the sky make them valuable assets for the exploration of ionospheric dynamics from very fine to medium size scales.\par
Consequently, radio arrays have been used, to a somewhat limited degree, for ionospheric studies.  Because of the relative stability of its electronics, quiet radio frequency interference (RFI) environment, and available VHF system (bands at 74 and 330 MHz), the Very Large Array (VLA) has been used almost exclusively in these efforts.  Seminal experiments were performed by \citet{jac92a,jac92b} using 330 MHz VLA observations of several sources to explore the environment of ionospheric waves above the VLA.  Among other results, they discovered a new class of magnetic eastward directed waves that were later shown to actually be located within the plasmasphere \citep{hoo97}.  Subsequent larger-scale investigations using data from the VLA Low-frequency Sky Survey \citep[VLSS;][]{coh07}, a survey of the northern sky at 74 MHz, have shown that the median behavior of ionospheric fluctuations observed by the VLA over several years is essentially consistent with turbulence \citep{coh09}.  Joint observations made with an all-sky optical camera and the VLA at 74 and 330 MHz by \cite{cok09} demonstrated the VLA's ability to aid in the exploration of the interaction between gravity waves generated lower in the atmosphere and relatively small-scale phenomena such as sporadic-$E$.  Recently, from a campaign using the VLA and the COSMIC satellite, \citet{dym11} were able to detect a relatively rare instance of a southeastward-propagating traveling ionospheric disturbance (TID).\par
In a companion paper \citep{hel11}, we described in detail the methods for extracting ionspheric information from the calibration of VLA VHF system data.  We demonstrated that when observing a bright cosmic source, the VLA VHF system was capable of achieving a typical $\delta \mbox{TEC}$ precision of $3 \times 10^{-4}$ TECU.  Since arrays such as the VLA are essentially only sensitive to the TEC gradient, we also detailed techniques for measuring time series of the TEC gradient to a precision of about $2 \times 10^{-4}$ TECU km$^{-1}$.  Here, we seek to develop this effort further by presenting new spectral analysis techniques for these TEC gradient measurements.  We will demonstrate that these techniques are capable of detecting and characterizing several phenomena from medium ($\sim\!100$ km) to small ($\sim\!5$ km) scales while also providing a broader statistical description of the spectrum of TEC fluctuations observable with the VLA.

\section{Observations and TEC Gradient Measurements}
The data used in this analysis came from a roughly 12-hour VLA observation of one of the brightest cosmic radio sources, Cygnus A (or ``Cyg A''; also known as 3C405) from a latitude and longitude of $34^{\circ} \: 04' \: 43.497''$ N and $107^{\circ} \: 37' \: 05.819''$ W (VLA program number AK570).  The observations were conducted during the night of 12-13 August, 2003 simultaneously at 74 and 327 MHz, both operating in dual polarization.  The VLA was in its ``A'' configuration spanning a circle with a diameter of about 40 km.  The observations consisted of a 35-minute block of time, or ``scan,'' and a second 12-hour scan with a temporal sampling of 6.67 seconds.  There was a moderate amount of geomagnetic activity ($K_p$ index  $\approx \! 2$--4) and the amount of solar activity was fairly average ($F10.7\!=\!123$ SFU).\par
Measurements of the TEC gradient toward Cyg A are described in detail in \citet{hel11}.  In short, the ionosphere adds an extra phase term to the complex visibility measure for each pair of antennas which is proportional to the difference in the TEC along the two lines of sight, or $\delta \mbox{TEC}$.  Thus, the interferometer is sensitive only to TEC gradients and not absolute TEC.  However, the Y-shape of the array (northern, southeastern, and southwestern ``arms'') makes measuring and performing Fourier inversions of the TEC gradient difficult.  This is in contrast to the standardized procedure of inverting visibilities to produce images of cosmic sources.  This is because the visibilities are functions of the \emph{differential} antenna positions and the TEC gradient is a function of the \emph{actual} antenna positions projected through the ionosphere.  This is discussed in more detail by \citet{hel11}.\par
Because of the consequences of the Y-shape of the VLA, \citet{hel11} established two ad hoc approaches to measuring the TEC gradient.  The first method estimates the full two-dimensional gradient at each antenna using a second-order, two-dimensional polynomial (Taylor series) on each time step.  While insensitive to fine-scale structure significantly smaller than the array, \citet{hel11} showed that this method is effective in recovering the gradient associated with relatively large, long-period disturbances visible at dusk, dawn, and intermittently throughout the night.\par
The second method numerically computes the projection of the TEC gradient along each of the VLA arms at each antenna and time step.  While little directional information can be obtained with these measurements, \citet{hel11} showed that there are many significant fine-scale fluctuations, especially near midnight local time, that are detected with this method and missed/damped within the polynomial-based method.  In the following section, we will detail the techniques we have developed to spectrally analyze the TEC gradient time series produced by these two methods.

\section{Spectral Analysis}
\subsection{Method}
To identify individual disturbances or sets of disturbances that passed over the array, we 
have developed the following Fourier-based method which uses the polynomial-based TEC gradient measurements.  This method is partially based 
on that used by \citet{jac92a}, but expands upon it to include some allowance 
for wavefront distortions and multiple wave fronts observed simultaneously.\par
The basic concept is to use both the apparent motion of Cyg A and the movements of the observed ionospheric disturbances to increase the spatial coverage of the VLA.  In other words, we will essentially convert temporal baselines into spatial ones, sampling the observed phenomena with a relatively long, 40 km-wide ``strip.''  Since there are presumably several disturbances contributing to the TEC gradient an each time step with a range of speeds, the temporal sampling cannot be converted into the same set of spatial samples for all observed phenomena.  Instead, we must separate the time series for each antenna into different Fourier modes and then analyze how the properties of each mode vary across the array to estimate the wavelength, speed, and direction of the contributing pattern(s).\par
To achieve this, we first assumed that the TEC fluctuations could be 
approximated as the sum of many individual oscillating modes, each having the 
following form
\begin{equation}
\mbox{TEC}(t) = A(r_{\perp}) \mbox{exp} \left [ i \left ( - \vec{k} \cdot \vec{r} + 2 \pi \nu t \right ) \right ]
\end{equation}
where $\vec{k}$ and $\vec{r}$ are the wavenumber and position vectors, respectively, 
$A(r_{\perp})$ is a complex amplitude which only varies perpendicular to $\vec{k}$ 
(i.e., wavefront distortions), and the temporal frequency, $\nu$, is assumed to be 
$=v |\vec{k}|/(2 \pi)$ where $v$ is the wave speed.  
For a single Fourier mode with temporal frequency, $\nu$, the Fourier transforms of the $x$ (N-S) and $y$ (E-W) partial 
derivatives are given by
\begin{eqnarray}
F_{x}(\nu) \simeq \left [ i k_{\nu , x} + \frac{\partial}{\partial x} A_{\nu}(r_{\perp}) \right ] \mbox{e}^{- i \vec{k_{\nu}} \cdot \vec{r}} \\
F_{y}(\nu) \simeq \left [ i k_{\nu , y} + \frac{\partial}{\partial y} A_{\nu}(r_{\perp}) \right ] \mbox{e}^{- i \vec{k_{\nu}} \cdot \vec{r}}
\end{eqnarray}
\par
From these equations, one can see that along a line parallel to $\vec{k}$, the $x$ and 
$y$ dependences of the phases of $F_x$ and $F_y$ are identical since both 
partial derivatives of $A_{\nu}(r_{\perp})$ are constant along this line.  There 
may be other lines in the $x,y$ plane where this is true.  For instance, in 
the absence of significant wavefront distortions, the $x$ and $y$ dependencies will 
be the same everywhere.  However, in the general case, the line parallel to 
$\vec{k}$ is the only one where these dependencies are required to be equal according 
to equations (2) and (3).  We have exploited this fact in 
our analysis to help determine the direction of $\vec{k}$ for different Fourier modes 
and consequently, the wavenumbers and velocities.\par
\begin{figure}
\noindent\includegraphics[width=6in]{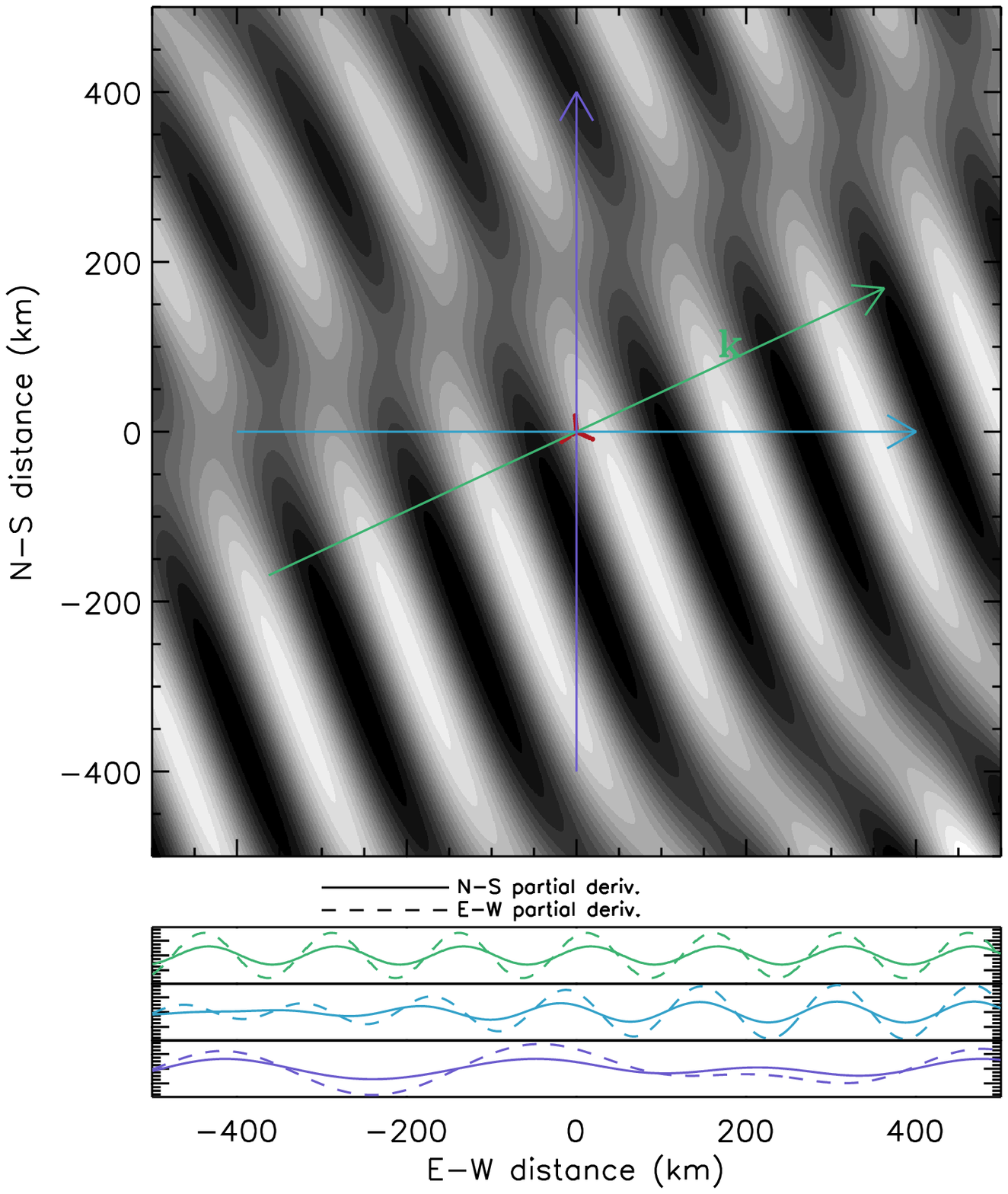}
\caption{To illustrate the technique used to extract wave properties described in \S 3, an example of a wave, including wavefront distortions, that might be observed by the VLA (shown as a red ``Y'').  The lower panels show the N-S (solid line) and E-W (dashed line) partial derivatives along each of the vectors plotted in the upper panel.}
\label{exwave}
\end{figure}

To illustrate this better, we have displayed an example of a wave, including wavefront distortions, in Fig.\ \ref{exwave}.  If one imagines this wave drifting over the VLA at a constant speed, one can see that the time series measured at two separate points will have identical periods, but will be out of phase.  For a constant speed, this phase difference will simply be proportional to the length of the separation between the points.  However, if one examines the N-S and E-W partial derivatives of this wave along different position vectors, one can see that the wavefront distortions add additional variations in phase.  We show this in the lower panels of Fig.\ \ref{exwave} where we have plotted the N-S and E-W partial derivatives, $F_x$ and $F_y$, along three different vectors shown on the image of the wave with color-coded arrows.  In all cases, $F_x$ and $F_y$ are out of phase with one another.  This phase difference is constant for the vector that is parallel to the wavenumber vector but varies with distance along the other two vectors.  Therefore, when wavefront distortions are present, the time series phases will have the same spatial derivative for both $F_x$ and $F_y$ only when points along the direction of the wave are considered.\par
While this derivation focuses on what one should expect for a single wave pattern, we must recognize that at a given temporal frequency, $\nu$, there will likely be contributions from several waves.  This can result from several factors including the finite size of the temporal window used for the Fourier inversion to the existence of phenomena such as turbulent cascades which have distributions of spectral power that span a large range of frequencies.  It this case, wave properties derived for a single mode using equations (2) and (3) will be weighted mean\footnote{Since the wave properties are derived from the phases of the combined Fourier transform of several phenomena, the derived wavenumbers and pattern speeds are not strictly weighted mean values but are proportional to the phases of the weighted mean Fourier transform of the phenomena.} values with more weight given to larger amplitude fluctuations.  In addition, even for a single dominant wave, the actual wavenumber and speed for a particular value of $\nu$ may change over the span of time used in the Fourier inversion.  Therefore, one should take the measured values for these properties to be averages over either the temporal window used for the Fourier transforms or the duration of the disturbance(s), whichever is shorter.\par
Thus, we may identify instances where single dominant waves are present by deriving wave properties using equations (2) and (3) for several temporal frequencies at each time step.  This can be done using Fourier transforms computed with a ``sliding'' window, one that is relatively wide but centered on each individual time step.  The derived spectral power, direction, and speed can then be examined as functions of time and $\nu$.  Any relatively large regions in the $t$,$\nu$ plane with relatively large power and uniform direction and speed are likely instances of dominant waves.  The wave properties measured in all other locations in the $t$,$\nu$ plane are likely composite values from several waves and can be used to study wave properties on a statistical basis.
\subsection{Derived Wave Properties}
Using the polynomial-based TEC gradient fits, we calculated $F_x$ and $F_y$ from equations (2) and (3) as functions of time by computing the 
discrete Fourier transform (DFT) of each polynomial coefficient within one-hour sliding 
windows for frequencies up to 12 hr$^{-1}$ (or, periods $>5$ minutes).  We chose 
the one-hour 
width because it was the same used to de-trend the $\delta \mbox{TEC}$ data 
\citep[see][]{hel11}.  Because of the one-hour box width, we excluded the first 35 minute 
scan from this analysis.  We also only computed $F_x$ and $F_y$ for times ranging 
from one half hour after the beginning of the second scan to one half hour before the 
end of the observing run so that a full hour of data could be used for each time 
step.\par
The DFT of each polynomial coefficient was then used to compute $F_x$ and 
$F_y$ at the locations of the antennas where the values of the polynomial fits 
are most reliable.  Upon inspection of this data, we found that the phases of 
both $F_x$ and $F_y$ at a single value of $\nu$ 
were typically well approximated with planes across the array. Because of this, 
at each time step and value of $\nu$, we fit planes to both these quantities.  
We then checked the combined (in quadrature) rms 
scatter about both of these fits against the rms difference between the phases 
of $F_x$ and $F_y$ to see if the $x$ and $y$ dependencies of these 
phases indeed differed across the array, indicating significant wavefront 
distortions.  In the vast majority ($>\!97$\%) of cases, the rms difference between the 
phases of the two partial derivative DFTs was larger than 
the combined rms scatter about the two planar fits (a median of four times larger).  
In these cases, we assumed that 
the azimuth angle (measured clockwise from north) of the line where the two planes 
intersected was the azimuth angle of $\vec{k}$ and that the derivative of the 
DFT phase along this line gave $| \vec{k} |$.  In the rare instance that there 
was no evidence of wavefront distortions, we computed a mean $x$ and $y$ slope 
for the DFT phases from the two planar fits and computed the direction and magnitude 
of $\vec{k}$ from these slopes.\par
\begin{figure}
\noindent\includegraphics[width=6in]{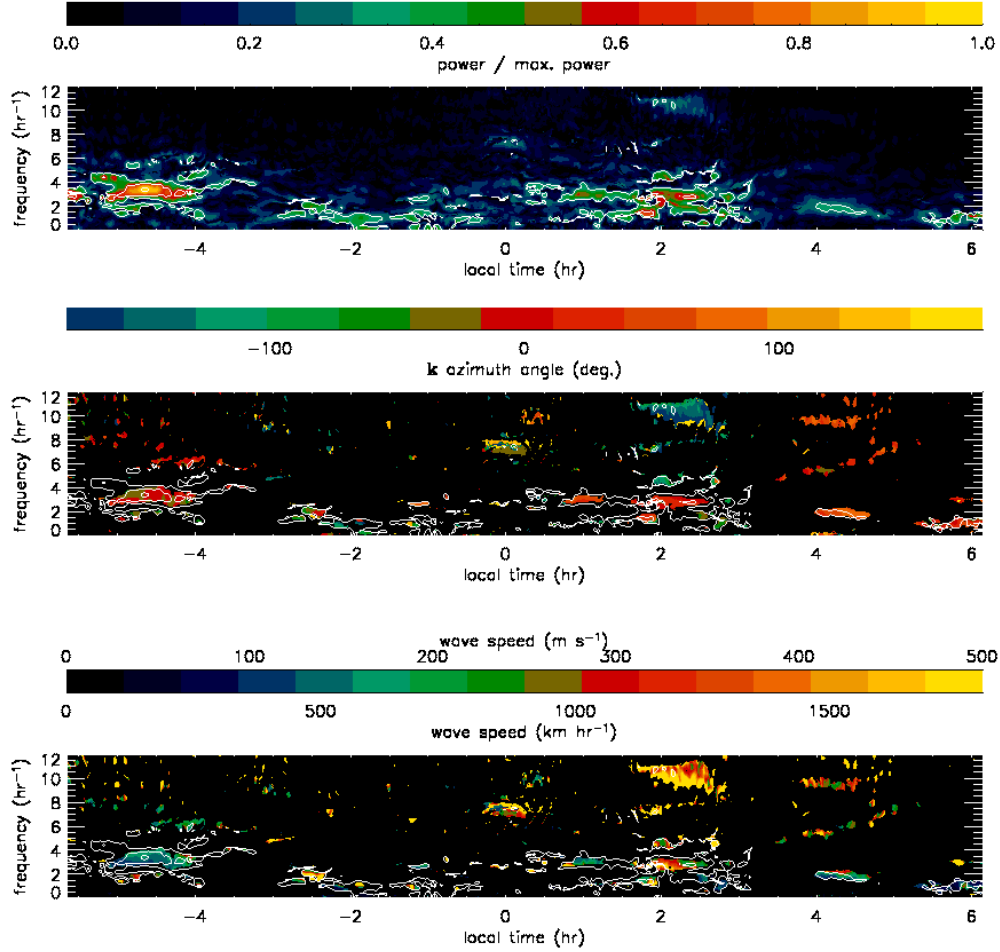}
\caption{Upper:  The power in the direction of the wavenumber vector, $\vec{k}$, 
as a function of local time 
(abscissa) and temporal frequency (ordinate) of Fourier modes within the $\delta \mbox{TEC}$ 
data.  Middle:  The azimuth angle of the wavenumber vector of each Fourier mode with power significantly larger than background levels (see \S 3.2).  Lower:  The velocity of each Fourier mode, using the same masking as in the middle panel.  In all panels, the white contours are of spectral power.}
\label{waveprop}
\end{figure}

The upper panel of Fig.\ \ref{waveprop} displays the total power of the TEC 
gradient in the direction of $\vec{k}$ as a function of local time and temporal 
frequency.  
While seemingly random fluctuations are frequently apparent, 
there are several instances where one can see significant 
detections of waves above the background.  Most of these waves are seen between 
frequencies of 1 and 5 hr$^{-1}$, or periods of 12 to 60 minutes.  This is roughly 
consistent with what is typical for MSTIDs.  To isolate these detections, we computed 
a mask by measuring the median and median absolute deviation (MAD) within elliptical ``annuli'' around each 
pixel that had dimensions of 1.5 hr in local time by 3 hr$^{-1}$ in $\nu$.  Any pixel with a 
value more than three times the MAD above the median for its annulus was considered 
a detection above the background.  We used this mask to display both the azimuth 
angle for $\vec{k}$ and the wave speed for significant 
detections in the middle and lower panels of Fig.\ \ref{waveprop}, respectively.\par
The plots in Fig.\ \ref{waveprop} imply that the candidate MSTIDs, i.e. fluctuations 
with periods $\sim 20$ minutes or greater, have estimated speeds between 100 and 
200 m s$^{-1}$, consistent with typical MSTIDs detected at night during northern 
hemisphere summer \citep{her06}.  
They also appear to move roughly westward/northwestward before midnight and toward the northeast after midnight.  While \citet{tsu07} showed that summer nighttime MSTIDs in the northern hemisphere predominantly move toward the southwest, \citet{her06} showed that near the west coast in California, they move almost due west and occasionally toward the northwest.  Thus, the waves observed before midnight are somewhat typical MSTIDs.  However, the northeastward directed waves seen after midnight are unusual and will be discussed in more detail below.

\subsection{Mean Power Spectra}
\begin{figure}
\noindent\includegraphics[width=6in]{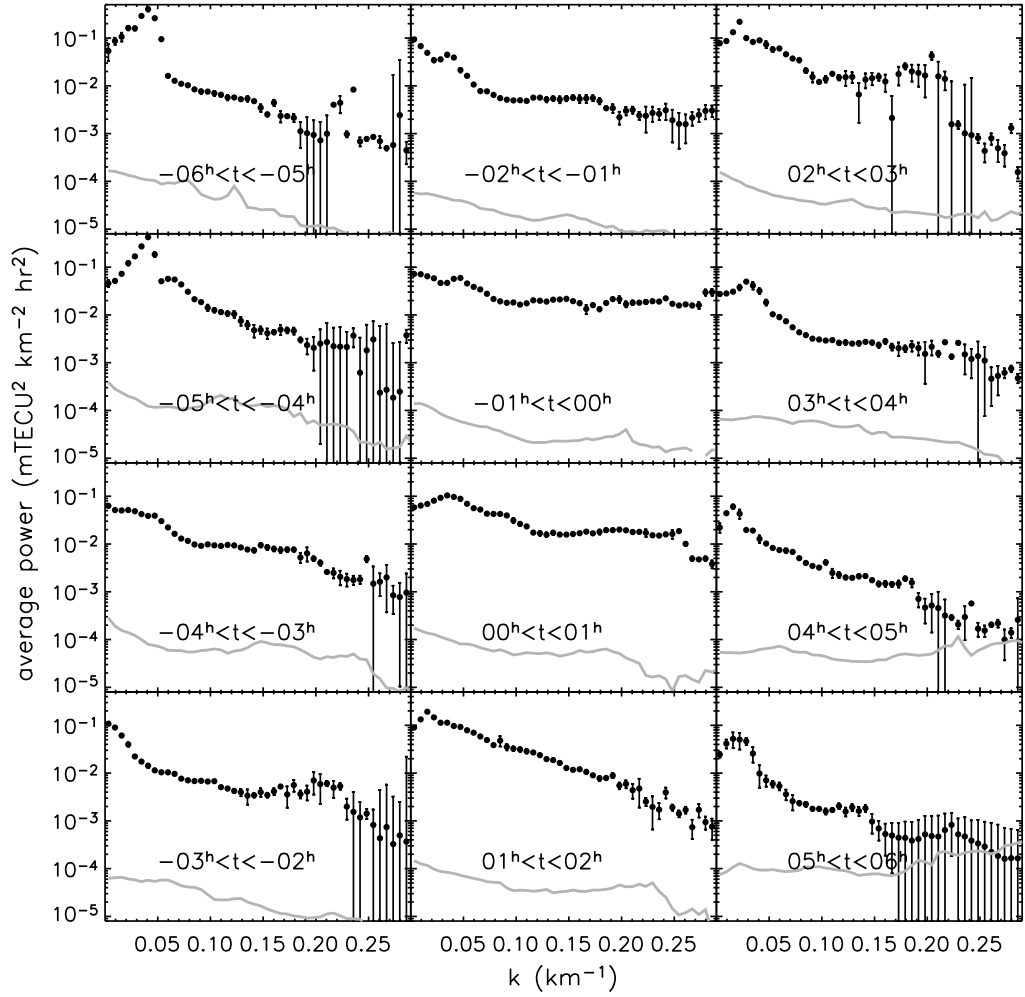}
\caption{Within one-hour bins, the mean power as a function of wavenumber, k, for the data 
displayed in Fig.\ \ref{waveprop}.  The noise-equivalent spectra (see \S 3.3) are also plotted 
as grey curves.}
\label{wavespec}
\end{figure}

While the results displayed in Fig.\ \ref{waveprop} are useful for identifying and examining instances of waves or groups of waves, the polynomial-based analysis can also be used to produce a more statistical description of the observed set of TEC fluctuations.  
In Fig.\ \ref{wavespec}, we have plotted the mean power within bins of 
wavenumber, $k$, for one-hour blocks of time.  We note that these wavenumbers 
have not been corrected for any Doppler shifts caused by the motion of 
the ionosphere pierce-points as the VLA tracked the apparent motion of Cyg A through the sky.  However, 
the results shown in Appendix A of \citet{hel11} demonstrated that the equivalent velocities 
at the estimated peak heights for the pierce-points are relatively small 
($\sim 20 \mbox{ m s}^{-1}$), except near the beginning and end of the observing 
run when Cyg A was rising and setting.  At these times, the main features 
detected in the data shown in Fig.\ \ref{waveprop} generally move roughly 
perpendicular to the pierce-point motions, implying that the Doppler shifts 
were relatively small during these times as well.\par
We have also plotted in Fig.\ \ref{wavespec} what 
we will refer to as ``noise-equivalent'' spectra.  These spectra were 
computed by performing the polynomial fits to each of the four 
$\delta \mbox{TEC}$ measurements (two bands and two polarizations) and 
computing the DFT of the difference between those fits and the fits to 
the final $\delta \mbox{TEC}$.  The wave analysis detailed above was then 
applied to these residual DFTs to estimate the results one would achieve 
if analyzing fluctuations that were simply noise.  The noise-equivalent 
spectra plotted in Fig.\ \ref{wavespec} are average spectra computed 
within each one-hour block and among the two bands and two polarizations.\par
These spectra imply that this technique can indeed detect significant 
power of structures as small as about half the size of the array (or, 
$k \approx 0.3 \mbox{ km}^{-1}$).  The MSTIDs visible in the upper panel 
of Fig.\ \ref{waveprop} are visible in most of the panels as significant 
bumps above the background at wavenumbers ranging from 0.015--0.05 km$^{-1}$, 
or wavelengths of about 130--500 km, which is again roughly consistent 
with the known properties of MSTIDs.  Few if any features are see at 
higher wavenumbers where the spectrum ranges from a simple power-law, 
reminiscent of turbulence, to relatively flat spectra, which may imply a 
population of intermediate-scale waves.\par
\begin{figure}
\noindent\includegraphics[width=6in]{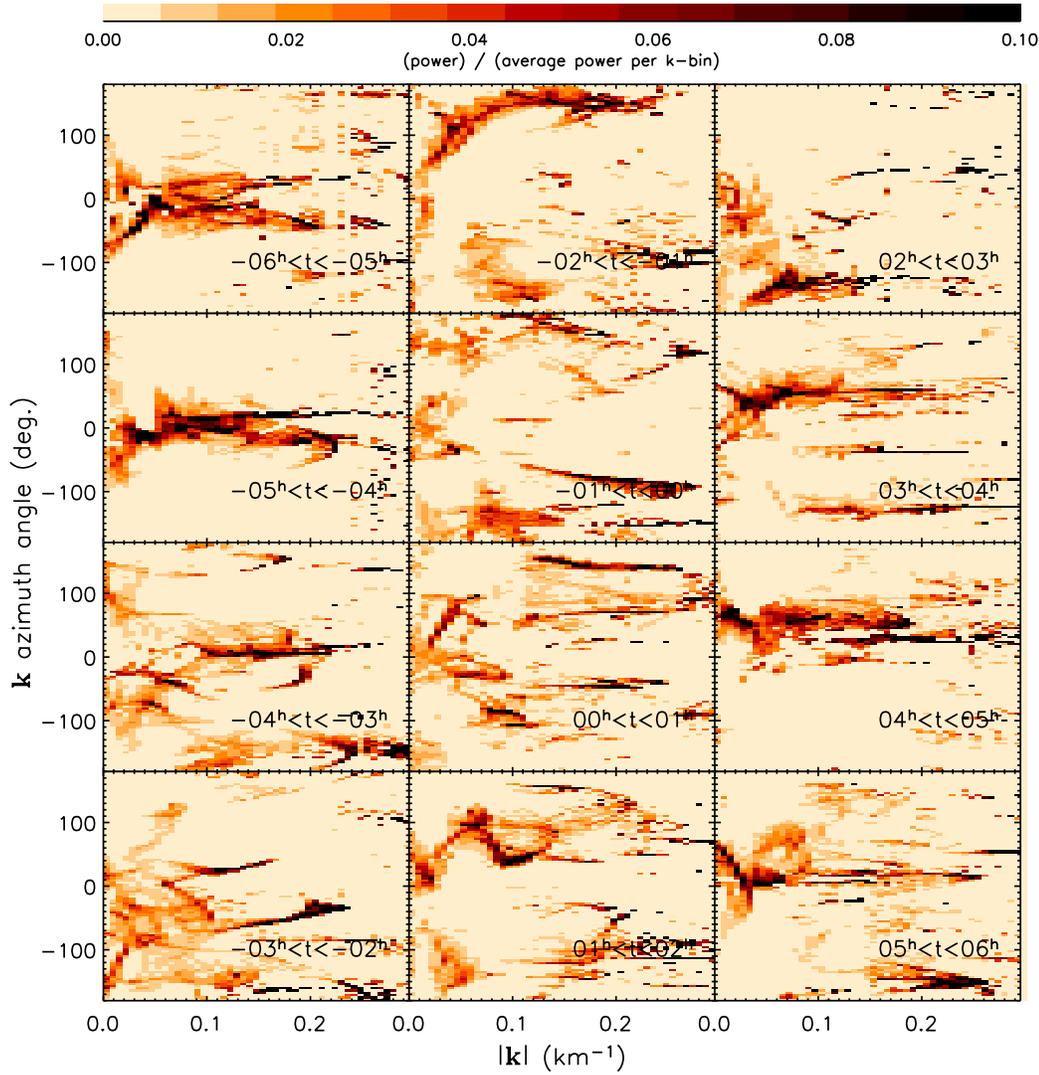}
\caption{Within one-hour bins, the mean power within bins of wavenumber, k, and azimuth angle.  The values within each bin have been normalized by the total power with all bins with the same wavenumber range to enhance the appearance of any detected features.}
\label{wavespecpa}
\end{figure}

To explore these data further, we have computed mean power spectra 
within bins of $\vec{k}$ azimuth angle.  We have displayed these as 
images in Fig.\ \ref{wavespecpa} where we have divided the data 
within the same wavenumber bin by the total power within that bin 
over all azimuth angles 
to enhance any detected features at higher wavenumbers.  
These images reveal that there are often intermediate-scale waves 
($k>0.1 \mbox{ km}^{-1}$, or wavelengths $<\!60$ km) detected moving in 
some preferred direction.  
Often, these waves appear to be moving in directions similar to 
detected MSTIDs.  This is especially apparent in the first three panels 
where one can see that even when the MSTIDs have disappeared by a 
local time of $-04^{\mbox{\scriptsize h}}$, the roughly northward moving 
intermediate-scale waves persisted.  The presence of strong MSTIDs do 
not appear to be required for these intermediate-scale waves as they 
seem to exist during all time periods.  However, the MSTIDs broadly 
appear to have the effect of ``focussing'' the intermediate-scale 
waves into a particular direction.\par
Those smaller-scale waves which are directed parallel to coincident MSTIDs may simply be small-scale structures within the MSTIDs themselves.  Indeed, our spectral analysis method has indicated that significant wavefront distortions are common within the VLA data.  However, the 
smaller waves which do not travel in the same direction as MSTIDs may be located within the $E$ region, most likely within sporadic-$E$ ($E_s$) layers given the prevalence of this phenomenon during summer nighttime.  In fact, \citet{cok09} showed that $E_s$ layers are likely the source of small-scale TEC fluctuations often observed with the VLA during summer nighttime.  In addition, $\sim\!10$-km-sized wave-like fluctuations have frequently been observed within $E_s$ layers as so-called quasi-periodic (QP) echoes \citep[e.g.,][]{fuk98,yam05}.  
If so, these waves may yield further insights into the coupling 
between the midlatitude nighttime $E$ and $F$ layers 
\citep[see][]{cos04a,cos04b} as the MSTIDs are likely located within the 
lower $F$ region \citep[see, e.g.,][]{her06}.  This will be discussed further in \S 5.

\subsection{Contemporaneous GPS and Ionosondes Data}
\subsubsection{GPS Data}
To partially validate our technique for detecting and characterizing waves, we 
have analyzed all available GPS data in the area for the time of our observations.  
From 12--13 August, 2003, there were six dual frequency GPS receivers operating within 
the state of New Mexico (station codes AZCN, NMSF, PIE1, SC01, TCUN, and WSMN) 
with publicly available data in RINEX format.  In contrast, there are currently more than 30 such stations in operation which bodes well for future studies conducted with the new VHF system being developed for the Expanded VLA (EVLA).  Data for each of the six stations operating in 2003 
were obtained and processed with standard GPS Toolkit \citep{tol04} 
routines (\texttt{DiscFix} and \texttt{ResCor}) 
to produce slant and vertical (for a height of 400 km) relative TEC values (i.e., 
no bias correction was done) for all available satellites.\par
For each 
station/satellite pair, the slant TEC data were de-trended according to 
\citet{her06} where they de-trended similar data by subtracting from each 
time step the average of two values, each an interval $\tau$ before or after 
the time step.  They noted that for a wave with a period $T$, this technique 
effectively multiplies the wave amplitude by a factor of 
$2 \: \mbox{sin}^2(\pi \tau / T)$.  They chose a value of $\tau = 300$s 
based on physical arguments of the expected periods of MSTIDs.  We have 
instead chosen $\tau = 600$s so that the de-trending technique was optimized 
for waves with periods similar to the MSTIDs detected within our 
VLA data.\par
After the GPS data were de-trended, they were corrected to 
vertical TEC values using the corrections computed by \texttt{ResCor} within GPS Toolkit.  
Within two-hour blocks, starting at a local time (relative to midnight, 13 August) 
of $-06^{\mbox{\scriptsize h}}$, 
we then identified station/satellite pairs which had relatively contiguous 
data ($>\!90$\% of the time interval covered) from one half hour before and one 
half hour after the time block.  We then computed the DFT of the de-trended 
TEC data within a one-hour 
sliding window to produce a spectrum at each time step within the two-hour 
block.\par
\begin{figure}
\noindent\includegraphics[width=6in]{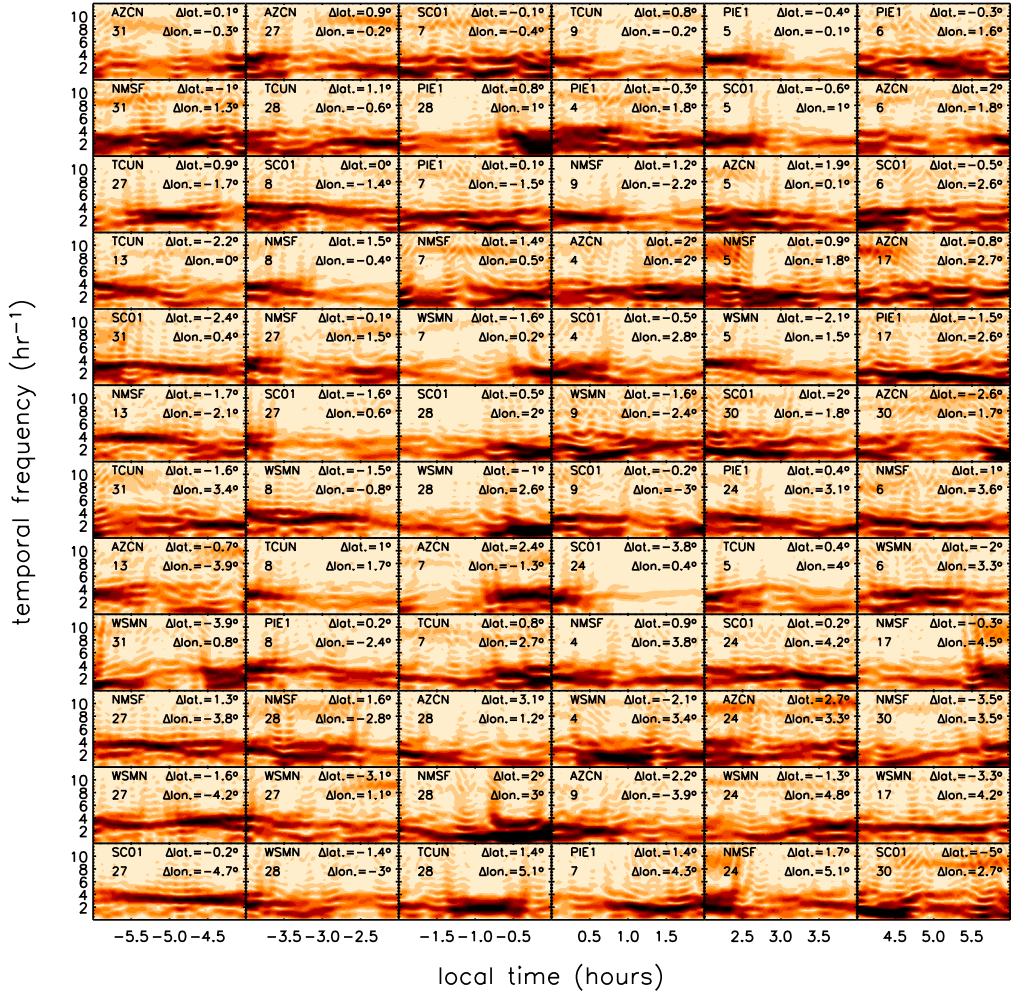}
\caption{Fluctuation spectra as a function of local time and temporal frequency for contemporaneous GPS observations (see \S 3.4.1).  Within each two-hour block, the spectra are sorted by the angular separation between their ionospheric pierce-points and that of Cyg A from top to bottom.  In each panel, the station code and GPS satellite number are printed in the upper left and the pierce-point latitude and longitude relative to that of the Cyg A pierce point is printed in the upper right.}
\label{exgps}
\end{figure}

These spectra are displayed in Fig.\ \ref{exgps} within each two-hour block for the 12 station/satellite pairs with ionospheric ``pierce-points'' closest to that associated with Cyg A.  Nearly all of these spectra show some level of wave activity with periods of about 12--60 minutes ($\nu\!=\!1$--5 hr$^{-1}$).  Those station/satellite pairs with pierce-points closest to that of Cyg A in particular show very good agreement with the VLA-derived spectrum displayed in the upper panel of Fig.\ \ref{waveprop}.  This is especially true for the top three rows of Fig.\ \ref{exgps} for local times of $-06^{\mbox{\scriptsize h}}$ to $-04^{\mbox{\scriptsize h}}$ and $01^{\mbox{\scriptsize h}}$ to $03^{\mbox{\scriptsize h}}$.\par
Unfortunately, the spacing among the GPS stations was too large to be 
able to track individual waves.  The closest separations among pierce-points 
is $>\!100$ km when separations on the order of tens of kilometers are required 
to overcome the effects of wavefront distortions \citep[e.g.,][]{her06}.  This makes it difficult to use the GPS data to 
directly validate the directional information we have determined for the 
VLA-detected waves displayed in the middle panel of Fig.\ \ref{waveprop}.  However, the fact that the spectra shown in Fig.\ \ref{waveprop} and \ref{exgps} observe many of the same waves is encouraging and at least partially validates our VLA-based approach.
\subsubsection{Ionosondes Data}
While there were no ionosondes operating in the immediate vicinity of the VLA, there were two relatively nearby at similar latitudes, one at Dyess Air Force Base (AFB) in west Texas ($32^{\circ} \: 30''$ N; $99^{\circ} \: 42'$ W), and one at Point Arguello in California ($35^{\circ} \: 36''$ N; $120^{\circ} \: 36'$ W).  The contemporaneous data from these stations are useful for interpreting our VLA observations.\par
For instance, northward or northeastward-directed MSTIDs, similar to those observed here at $01^{\mbox{\scriptsize h}}$--$03^{\mbox{\scriptsize h}}$ and $04^{\mbox{\scriptsize h}}$--$04^{\mbox{\scriptsize h}}30^{\mbox{\scriptsize m}}$, have been observed over Japan with the SuperDARN Hokkaido HF radar and an airglow imager by \citet{shi08}.  Like the northeastward-directed MSTIDs described here, these were detected at midlatitudes during nighttime.  \citet{shi08} noted that the roughly northward-directed MSTIDs were observed with the SuperDARN array most frequently in May and August.  They also noted that the northeastward-directed waves were coincident in time with a drop in the height of the F-region as indicated with nearby ionosondes data.\par
\begin{figure}
\noindent\includegraphics[width=6in]{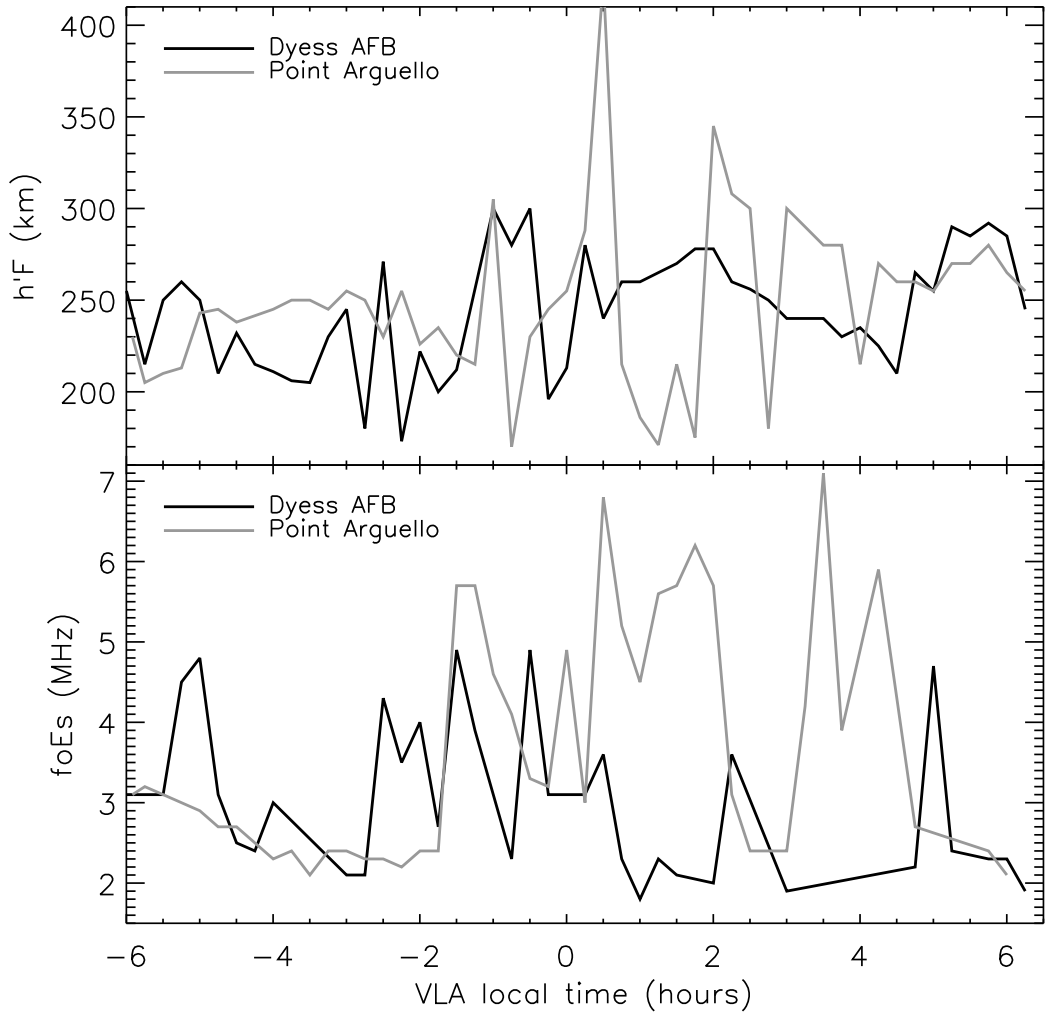}
\caption{For two nearby ionosondes stations, Dyess Air Force Base ($32^{\circ} \: 30''$ N; $99^{\circ} \: 42'$ W) and Point Arguello ($35^{\circ} \: 36''$ N; $120^{\circ} \: 36'$ W), h$^\prime$F (upper) and foEs (lower) as functions of VLA local time.}
\label{ionos}
\end{figure}

To demonstrate that this may also be the case for our VLA observations, we have plotted h$^\prime$F for both the Dyess AFB and Point Arguello ionosondes as functions of VLA local time in the upper panel of Fig.\ \ref{ionos}.  The Point Arguello data show a substantial drop in h$^\prime$F at around $01^{\mbox{\scriptsize h}}$ at the start of the first observed instance of northeastward directed waves.  The Dyess AFB data shows a decline in h$^\prime$F which begins around $02^{\mbox{\scriptsize h}}$ and reaches a minimum at about $04^{\mbox{\scriptsize h}}30^{\mbox{\scriptsize m}}$, corresponding to the middle of the second period of northeastward MSTID activity.  This indicates that it is plausible that we have observed a similar phenomenon with VLA over New Mexico as was found by \citet{shi08} over Japan.\par
The ionosondes data also serve to validate our claims that $E_s$ layers were present during our observations.  In the lower panel of Fig.\ \ref{ionos}, we have plotted foEs for both ionosondes stations.  Both stations observed E-region reflections with maximum frequencies between 2 and 6 MHz throughout the night, strongly indicating that $E_s$ layers were present.  The most active time for the appearance of these layers seems to be between $-02^{\mbox{\scriptsize h}}30^{\mbox{\scriptsize m}}$ and $04^{\mbox{\scriptsize h}}30^{\mbox{\scriptsize m}}$ VLA local time, especially for the Point Arguello station.  As evidenced by Fig.\ \ref{wavespec} and \ref{wavespecpa}, relatively small-scale fluctuations are visible within the VLA data throughout this time period.  They are especially prominent from $-02^{\mbox{\scriptsize h}}$ to $01^{\mbox{\scriptsize h}}$ where their presence has made the mean spectra shown in Fig.\ \ref{wavespec} virtually flat for wavenumbers $>\!0.13$ km$^{-1}$.

\section{Statistical Wave Properties}
\subsection{Arm-based Spectra}
While the Fourier analysis detailed in \S 3 has yielded interesting results, 
it neglects the ability of the VLA in its A configuration to detect 
fluctuations on scales as small as a few kilometers.  To make full use 
of the data, we have developed a complementary spectral analysis which 
uses the projected TEC gradients measured along each arm which are sensitive to fluctuations with sizes as small as the shortest antenna spacings within the VLA.\par
To analyze the projected gradients in a manner similar to the method used for the polynomial-based gradients, we first performed fast Fourier transforms (FFTs) of the times series for each antenna.  This was done within a sliding window 
of approximately one hour (512 time steps, or 56.9 minutes) centered on each time step.  Then, for a single arm, time step, and temporal frequency, 
we unwrapped the 
phases of the FFTs along the arm and numerically computed the 
derivatives of these phases with respect to distance along the arm
 to obtain estimates of the (mean) projected wavenumber, $k_{proj}$, of each Fourier mode.  
As noted in \S 3, our Fourier analysis of the polynomial fits revealed 
that significant wavefront distortions were common.  Because of this, 
we expect that these $k_{proj}$ estimates are not reliable 
for tracking individual disturbances along any of the arms.  However, since the effects of 
such distortions should average to zero over time, these estimates 
are sufficient to perform a statistical analysis of the overall 
population of waves seen by the VLA from a few to hundreds of 
kilometers in wavelength.\par
\begin{figure}
\noindent\includegraphics[width=6in]{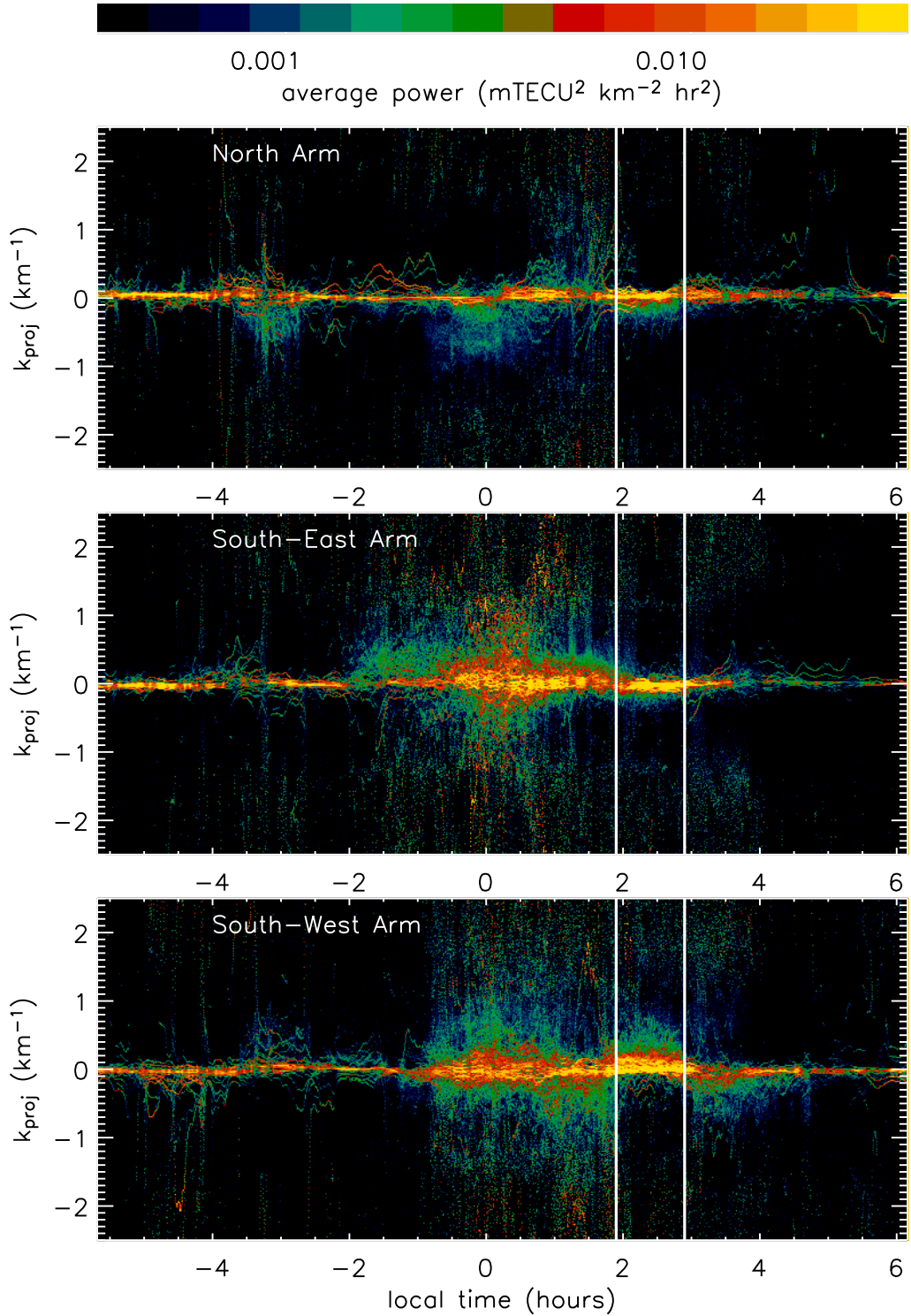}
\caption{For each of the VLA arms, the total power as a function of projected wavenumber, 
$k_{proj}$, within one-minute bins.  The region highlighted with vertical white lines in each panel is discussed in \S 4.1.}
\label{armspec}
\end{figure}

To begin such a statistical analysis, we binned the FFT and $k_{proj}$ 
data within one-minute intervals and computed the mean FFT power within 
bins of $k_{proj}$ for each arm.  The resulting spectra are displayed 
in Fig.\ \ref{armspec} where positive projected wavenumbers correspond to directions outward from the center of the array, or ``up'' the arm and 
negative projected wavenumbers correspond to inward directions, or ``down'' the arm.  
Since these are projected wavenumbers, instances where there are 
groups of waves moving in a particular direction will show up in 
these plots as over-densities in the positive (negative) direction 
for one arm, and in the negative (positive) direction for the other 
two.  In cases where a group of waves moves nearly perpendicular to 
a particular arm, they will not show up at all in that arm's spectrum.  
A good example of this can be seen between roughly $02^{\mbox{\scriptsize h}}$ 
and $03^{\mbox{\scriptsize h}}$ local time (highlighted with vertical white lines in the panels of Fig.\ \ref{armspec}).  Here, 
there is an obvious over-density on the positive side of the spectrum 
for the southwestern arm and mild over-densities on the negative sides 
of the spectra for the northern and southeastern arms.  This implies 
that these fluctuations are moving almost directly toward the southwest.  
Part of this group of waves can also be seen the panel of the upper 
right corner of Fig.\ \ref{wavespecpa}, but in Fig.\ \ref{armspec}, 
they can be seen 
to extend to wavenumbers of at least 0.7 km$^{-1}$, or wavelengths 
at least as small as 9 km.\par
\begin{figure}
\noindent\includegraphics[width=6in]{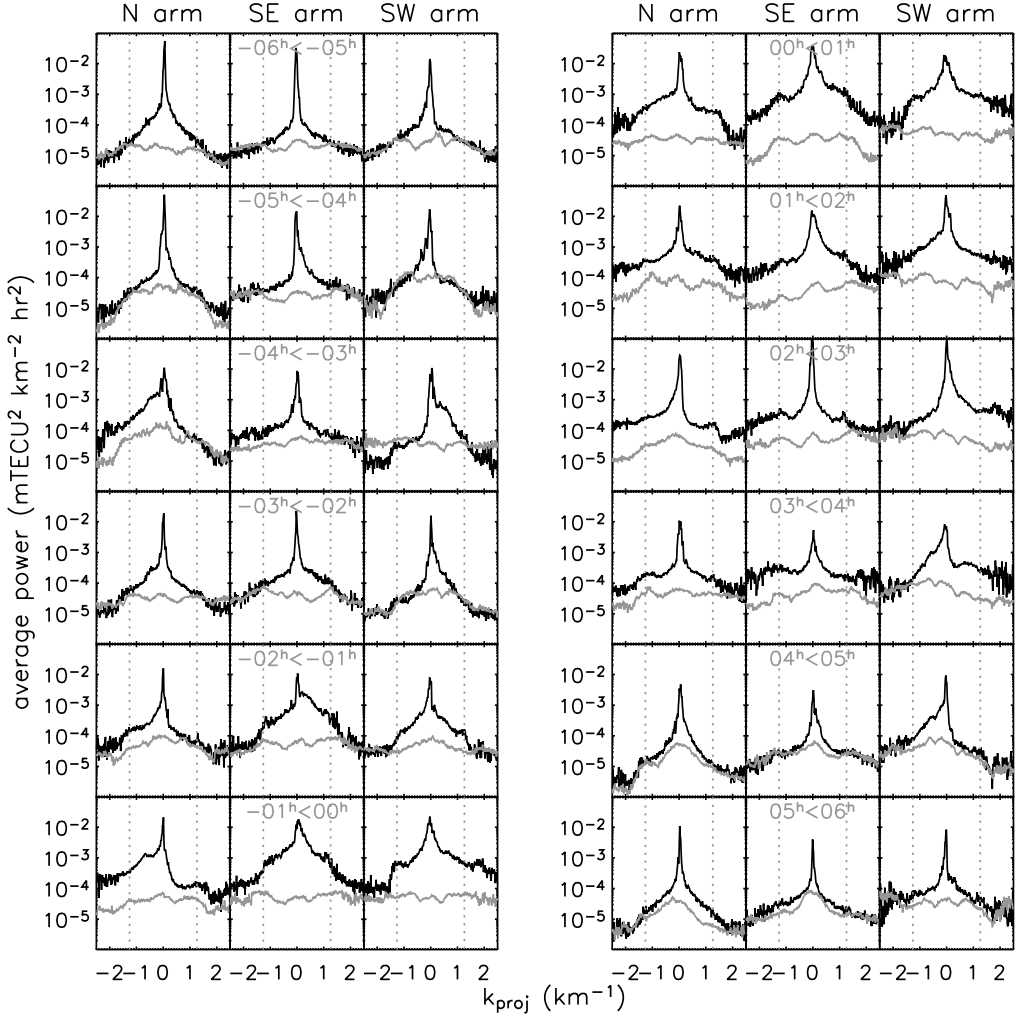}
\caption{Within one-hour bins, the mean power within bins of projected wavenumber, $k_{proj}$, 
for each of the VLA arms.  The noise-equivalent spectrum (see \S 4.1) is also plotted in 
each panel as a grey curve.  The vertical dotted grey lines denote values of $k_{proj} = \pm 1.26 \mbox{ km}^{-1}$ which corresponds to the approximate Nyquist sampling limit for the average antenna separation of 2.5 km along the VLA arms (in A configuration).}
\label{armspec1hr}
\end{figure}

To establish the minimum scales on which this analysis can be 
considered reliable, we have also computed noise-equivalent spectra 
using the full two-band, two-polarization data as we did with the 
polynomial-based Fourier analysis in \S 3.  We have plotted in 
Fig.\ \ref{armspec1hr} mean spectra for each arm within one hour 
blocks of time along with the corresponding noise-equivalent 
spectra.  In some instances, the spectra are always above the 
noise-equivalent spectra.  However, in general, the spectra reach 
the noise level at projected wavelengths of about 5 km, or 
$k_{proj}= \pm 1.26 \mbox{ km}^{-1}$ (see the vertical dotted lines in 
Fig.\ \ref{armspec1hr}).  This is in keeping with the mean antenna 
separation along the arms of about 2.5 km, implying an approximate 
Nyquist sampling limit of 5 km.  Because of this, we consider these 
spectra most reliable for $|k_{proj}|\!<\!1.26 \mbox{ km}^{-1}$.\par
We note that for most of these one-hour mean spectra, the relatively small-scale ($|k_{proj}|\!>0.1$) portions of the spectra that are above the noise level are roughly consistent with what is expected for $E_s$ layers.  From the data presented by \citet{cok09}, a typical time series of $E_s$ activity would yield values for the spectral power ranging from about $10^{-4}$ up to $10^{-3}$ mTECU$^2$ km$^{-2}$ hr$^2$.  This is nearly exactly the range inhabited by the intermediate/small-scale regions of the one-hour mean spectra plotted in Fig.\ \ref{armspec1hr}.  In addition, the contemporaneous ionosondes data presented in \S 3.3.2 and Fig.\ \ref{ionos} show that $E_s$ layers were likely present during the VLA observations.\par
For the remainder of the analysis, we found it useful to define 
three size classes for the detected waves: (1) medium scale, 
$|k_{proj}|<0.1 \mbox{ km}^{-1}$ which includes the full range of 
MSTID sizes, (2) intermediate scale, $0.1<|k_{proj}|<0.3 \mbox{ km}^{-1}$, 
which represents the remaining range of scales one can probe with 
our previous polynomial-based approach, and (3) small scale, 
$0.3<|k_{proj}|<1.26 \mbox{ km}^{-1}$ representing the scales that 
cannot be probed with the polynomial fits.\par
\begin{figure}
\noindent\includegraphics[width=6in]{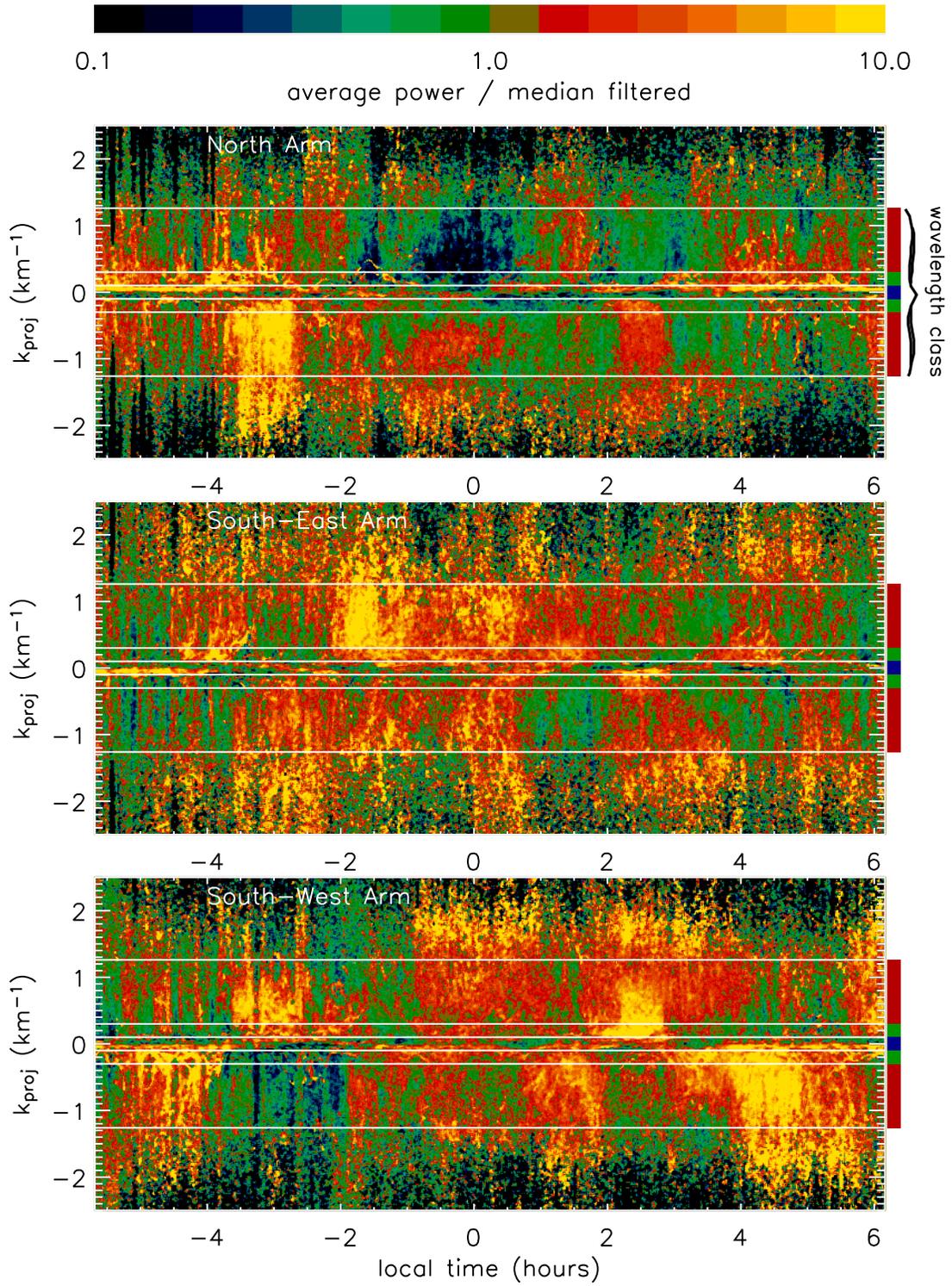}
\caption{The spectra from Fig.\ \ref{armspec} normalized by the estimated background spectrum and with the three wavenumber classes highlighted (see \S 4.1).}
\label{armspecmed}
\end{figure}

We have re-displayed the arm-based spectra in Fig.\ \ref{armspecmed} 
with these three size classes marked.  In this representation, we 
have divided the spectra by an estimate of the spectra of background 
fluctuations, i.e., those that do not move in a preferred direction.  
We have estimated this background at each value of $|k_{proj}|$ and time step using a combination of the spectra from all three arms.  In this process, only one of the two directions, up or down the arm, was used for each arm.  In each case, the direction with the lowest power was used as it was more likely to be dominated by isotropic fluctuations.  The median power among all three arms (within $\pm 1$ pixel in both time and $|k_{proj}|$) was then taken to be the value for the background.  For displaying purposes only, we also applied a three-pixel 
square median filter to the arm-based spectra and divided these by 
the estimated background spectra and have shown them in Fig.\ 
\ref{armspecmed}.  In this representation, one can see that groups of 
waves at virtually all detectable scales are seen moving in a variety 
of preferred directions at different times.

\subsection{Group Wave Properties}
To exploit the arm-based spectra described above and displayed in Fig.\ 
\ref{armspec}--\ref{armspecmed} to identify groups of waves moving in 
a particular direction, we have developed the following method.  To 
begin, we computed a weighted mean value of $k_{proj}$ for each size 
class, arm, and one-minute interval using the spectral power as the weight.  
If there is a significant population of waves within any of the three 
classes at a particular time step moving in a similar direction, 
these weighted mean values will follow a cosine as a function of the 
azimuth angle of the arms which will peak in the direction of the true 
wavenumber vector, $\vec{k}$.\par
\begin{figure}
\noindent\includegraphics[width=6in]{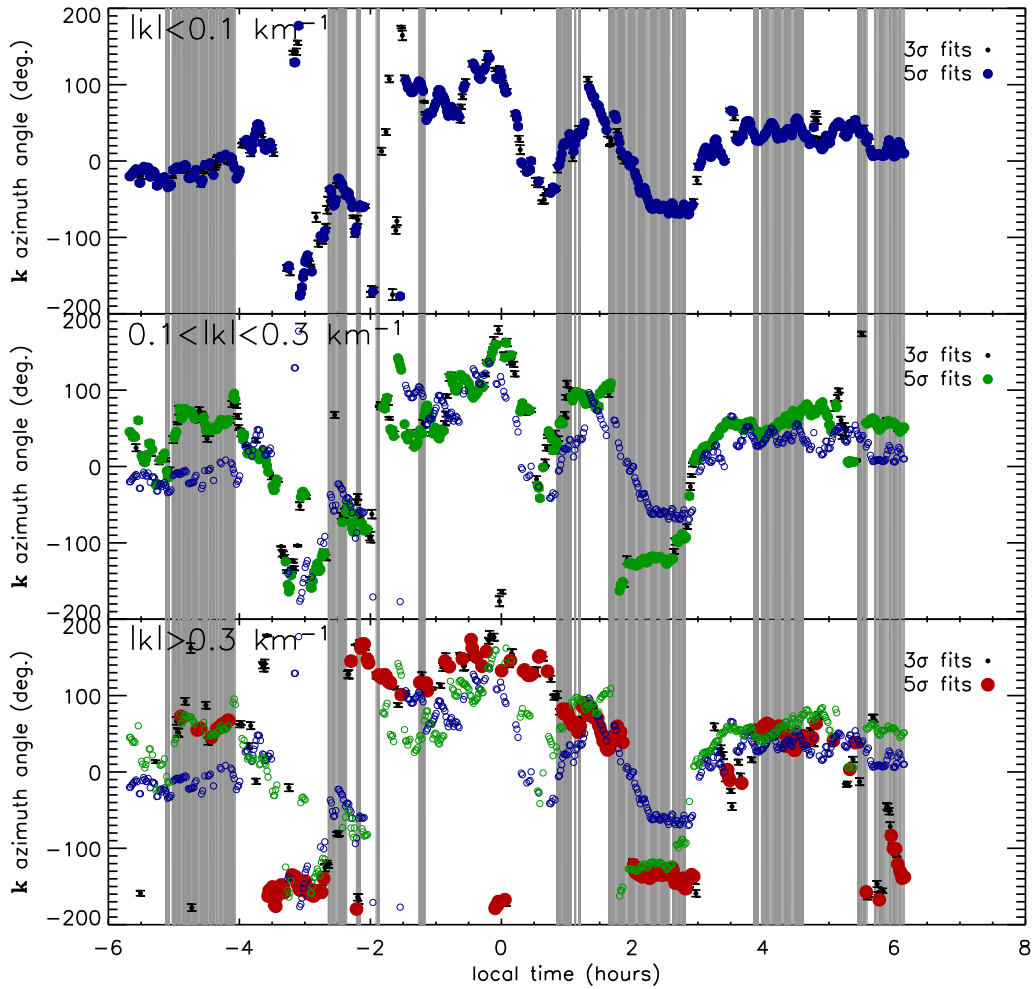}
\caption{The wavenumber vector azimuth angle for $3\sigma$ and $5\sigma$ detections (see \S 4.2) for each of the three wavenumber classes.  Time ranges where MSTIDs 
were detected from the data displayed in the upper panel of Fig.\ \ref{waveprop} are shaded 
in grey.}
\label{armspecpa}
\end{figure}

To detect such groups of waves, we have 
fit a simple cosine model to the weighted mean $k_{proj}$ values 
at each time step, including the nearest two time steps so that the 
fit would be better constrained and so some level of temporal 
coherence would be imposed.  We then compared the amplitude of the 
fitted cosine with the rms scatter about the fit among the nine 
values used (i.e., three arms and three time steps) to assess the 
significance of the detection of a group of waves.  We have plotted 
the azimuth angles for $\vec{k}$ for both $3\sigma$ and $5\sigma$ 
detections for each of the three size classes in the panels of 
Fig.\ \ref{armspecpa}.  The $5\sigma$ detections for the medium 
and intermediate scale waves have been re-plotted in the panels 
of the smaller size class(es) for comparison.  For reference, 
we have also shaded in grey time ranges where MSTIDs where 
detected using the polynomial-based technique as described in \S 3.\par
In general, the $\vec{k}$ directions for the medium-scale class 
agree with those seen for the detected MSTIDs displayed in Fig.\ 
\ref{waveprop}, which serves to partially validate our arm-based 
method.  Most of the time, the azimuth angles for the intermediate 
and small classes roughly agree, indicating that they are likely 
part of the same distribution of waves.  Both often agree with the 
azimuth angles determined for the medium class, but are at times 
significantly different, most notably during two instances where 
MSTIDs were detected.\par
We have also determined approximate speeds for all of the detected 
groups of waves which we have plotted in Fig.\ \ref{armspeckv} 
with their values of 
$|\vec{k}|$ taken from the above described cosine fits.  The speeds 
were computed using the mean temporal frequency within bins of 
$k_{proj}$ that were within $\pm 0.1$ km$^{-1}$ for the small class 
and $\pm 0.05$ km$^{-1}$ for the intermediate and medium classes 
of the cosine fit-determined value of $|\vec{k}|$ from each arm.  A spline fit 
to the data as a function of arm azimuth angle was then used to 
estimate the mean temporal frequency along the direction of $\vec{k}$ 
where waves, rather than background fluctuations, are more likely 
to dominate.  The speed was then 
computed using this frequency and the value of $|\vec{k}|$ from the 
cosine fit.  The speeds agree in general with what can be seen in 
Fig.\ \ref{armspecpa}, that the medium-scale class of waves are a 
somewhat distinct population in terms of direction and speed whereas 
the intermediate and small scale classes tend to have similar 
directions and speeds.
\subsection{Small-scale Waves}
\begin{figure}
\noindent\includegraphics[width=6in]{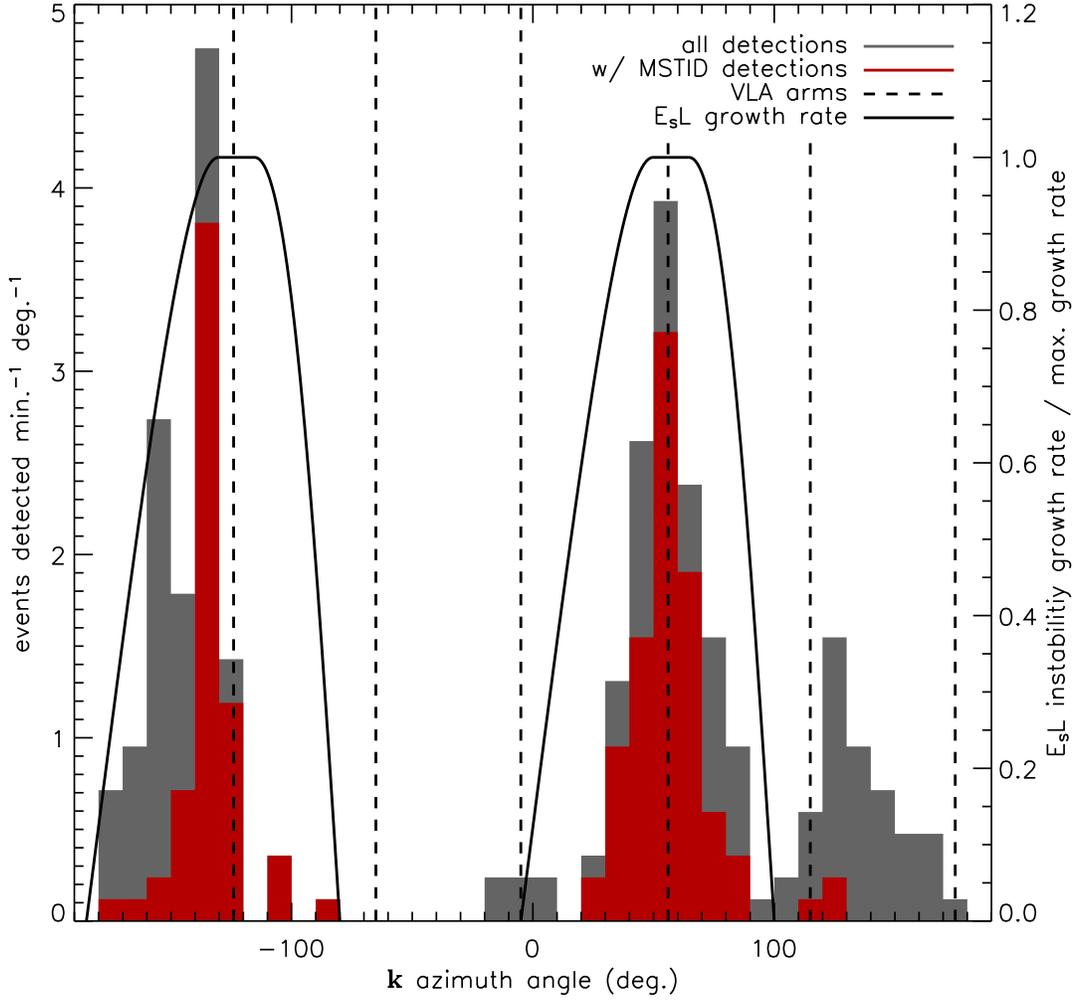}
\caption{The distribution of $\vec{k}$ azimuth angles for $5 \sigma$ detections of small-scale moving structures displayed in Fig.\ \ref{armspecpa} (shaded grey histograms).  The same distribution for those detections that were coincident with MSTID detections (see Fig.\ \ref{armspecpa}) are displayed in red.  The growth rate for the $E_{\mbox{\scriptsize s}}$ layer instability is plotted as a black curve (see right ordinate for scale), allowing for a range in meridional wind components \citep[see \S 4.3 and][]{cos02}.}
\label{padist}
\end{figure}

To further explore the nature of the small-scale waves in particular, 
we have plotted the distribution of azimuth angles for the $5\sigma$ 
detections in Fig.\ \ref{padist} for the small-scale class only.  
From this it can be seen that the distribution has four distinct 
peaks at angles of roughly $-155^{\circ}$, $-135^{\circ}$, $55^{\circ}$, 
and $120^{\circ}$.  For reference, we have also plotted the azimuth angles of the three VLA arms (both up and down the arms) as vertical dashed lines.  From this, one can see that the group of waves with a peak azimuth angle of $55^{\circ}$ were moving almost directly up the southwestern arm.  This indicates that this group of waves is likely small in extent as well as in wavelength.  This is because for groups of waves substantially smaller than the array, the arm-based method we described above will be biassed toward waves traveling along one of the three arms because the number of antennas that ``see'' these waves will be maximized in this case.\par
In contrast, groups of waves which span all or nearly all of the array will essentially be seen by all the antennas for any direction and will experience no such bias.  This appears to be the case for the group of waves moving roughly southeastward and for the waves with a peak azimuth angle of $-155^{\circ}$.  The peak of the distribution for the southeastward-directed waves is somewhat skewed toward the azimuth of the southeastern arm.  However, the bulk of these waves have azimuth angles between that of the southeastern (up) and northern (down) arms, indicating that any influence of the arm-based bias on these waves is minimal.  The group of waves moving toward an angle of $-135^{\circ}$ peak near but not at the azimuth angle of the southwestern arm, indicating that their measured distribution has been influenced by the arm-based bias, but to a lesser degree than those whose distribution peaks near $55^{\circ}$.\par
Because of this bias, the paucity of waves directed either due west or east only indicates that there were few groups of small-wavelength waves moving in these directions which also spanned the array.  However, the lack of small-scale waves seen directed northwest, due north, or due south (i.e., up the southeastern arm or up/down the northern arm) indicates a genuine absence of such phenomena and demonstrates that the population of small-scale waves was far from isotropic.\par
In terms of the implied orientation of the wavefronts, 
the peaks at $-135^{\circ}$ and $55^{\circ}$ are essentially the same 
since they are roughly $180^{\circ}$ apart.  They are both conspicuously 
close to the required orientation for the $E_{s}$ layer instability
 (or, ``$E_s$LI'') described in detail by \citet{cos02}.  They 
demonstrated that an $E_s$ layer is unstable against perturbations 
with wavefronts aligned from northwest to southeast.  In particular, 
for no meridional wind component, the optimum orientation is 35$^{\circ}$ 
west of magnetic north.  For the VLA, where the magnetic declination 
is about 10$^{\circ}$, this corresponds to $\vec{k}$ azimuth angles 
of either $-115^{\circ}$ or $65^{\circ}$.  If there is a significant 
meridional wind component, the optimum angle can be as much as 
$15^{\circ}$ closer to due south, giving a range of optimum position 
angles of $-130^{\circ}$ to $-115^{\circ}$ or $50^{\circ}$ to $65^{\circ}$.  
This is roughly consistent with the locations of the two largest 
peaks in Fig.\ \ref{padist}.  However, the exact peak azimuth angles of these two distributions are not precisely constrained given the bias toward the direction of the VLA arms described above.  We can only say that they are generally moving either northeast or southwest.\par
To illustrate the agreement with the predictions of the $E_s$LI model, we have 
computed the azimuth angle (at the VLA) dependence of the $E_s$LI 
assuming a magnetic dip angle of $45^{\circ}$ according to \citet{cos02}, 
including a range of peak azimuth angles of $15^{\circ}$.  We have 
plotted a scaled version of the growth rate in Fig.\ \ref{padist}.  From this 
curve, one can see that the bulk of the two groups of waves with 
azimuth angles near the optimum values for the $E_s$LI are contained 
within the regions that are within $\sim 1/2$ of the maximum growth 
rate.\par
If we consider only those waves located in time ranges where MSTIDs 
were detected, the agreement with the predictions of the $E_s$LI growth 
rate are even better (see the red histograms in Fig.\ \ref{padist}).  In 
fact, only a small fraction of such waves ($\sim\! 2$\%) have azimuth 
angles where the $E_s$LI growth rate is $\leq\! 0$.  In contrast, for 
all waves, about 16\% lie outside this region.  One can see from the red histograms in Fig.\ \ref{padist} that the azimuth angle distribution for the southwest-directed waves coincident with MSTIDs is significantly skewed with a tail extending away from the azimuth angle of the southwestern VLA arm.  This indicates that the arm-based bias discussed above has a larger effect on these waves than the other southwest-directed waves which are not seen with MSTIDs and have a more symmetric distribution.  This further implies that these groups of waves and their northeast-directed counterparts are relatively small in extent as compared to the other detected groups of small-scale waves.\par
We note that small-scale waves were detected during nearly all observed instances of MSTIDs.  This indicates that the preference for small-scale waves aligned northwest to southeast during MSTID activity is not simply a product of the arm-based bias.  In other words, there is no indication that there were periods of MSTID activity where small-scale waves were not detected because they were not directed along a VLA arm and their extent was too small to be detected with the arm-based method.  This preference for northwest to southeast aligned, small-scale waves is again consistent 
with the notion that these waves are associated with the $E_s$LI.  This is because 
according to \citet{cos04a}, the coupling between the Perkins instability 
in the $F$ region and the 
$E_s$LI is itself unstable.\par
The growth rate for the coupled instability 
is maximized for 
instances where both the $F$ and $E_s$ layer perturbations are aligned 
from northwest to southeast.  From Fig.\ \ref{armspecpa}, we can see that 
while the small-scale waves are so aligned, the MSTIDs are not necessarily.  
However, we note that the range of allowed orientations for the Perkins 
instability is much broader than that for the $E_s$LI as it depends 
somewhat on the strength and orientation of the $F$ region neutral wind.\par
We also note that models of this instability \citep{yok09,yok10} have 
shown that random perturbations within an $E_s$ layer are enough to 
form MSTIDs in the lower/bottom part of the $F$ region through the $E$-$F$ coupling 
mechanism.  The MSTIDs can then, through this same coupling, help 
northwest to southeast aligned perturbations grow within the $E_s$ 
layer.  \citet{yok09} confirmed that this process works most 
efficiently when conditions are right in the lower/bottom part of the $F$ region to 
form northwest to southeast aligned MSTIDs.  However, they also 
showed that while weaker, MSTIDs of different orientations (in their 
example, aligned north to south) could be formed through this 
process.  In this case, the $E_s$ layer waves that formed were 
still aligned northwest to southeast due to the more rigid 
directional constraints of the $E_s$LI.  Therefore, the results 
show here are quite consistent with the predictions of $E$--$F$ 
coupling in the nighttime, midlatitude ionosphere.\par
\begin{figure}
\noindent\includegraphics[width=6in]{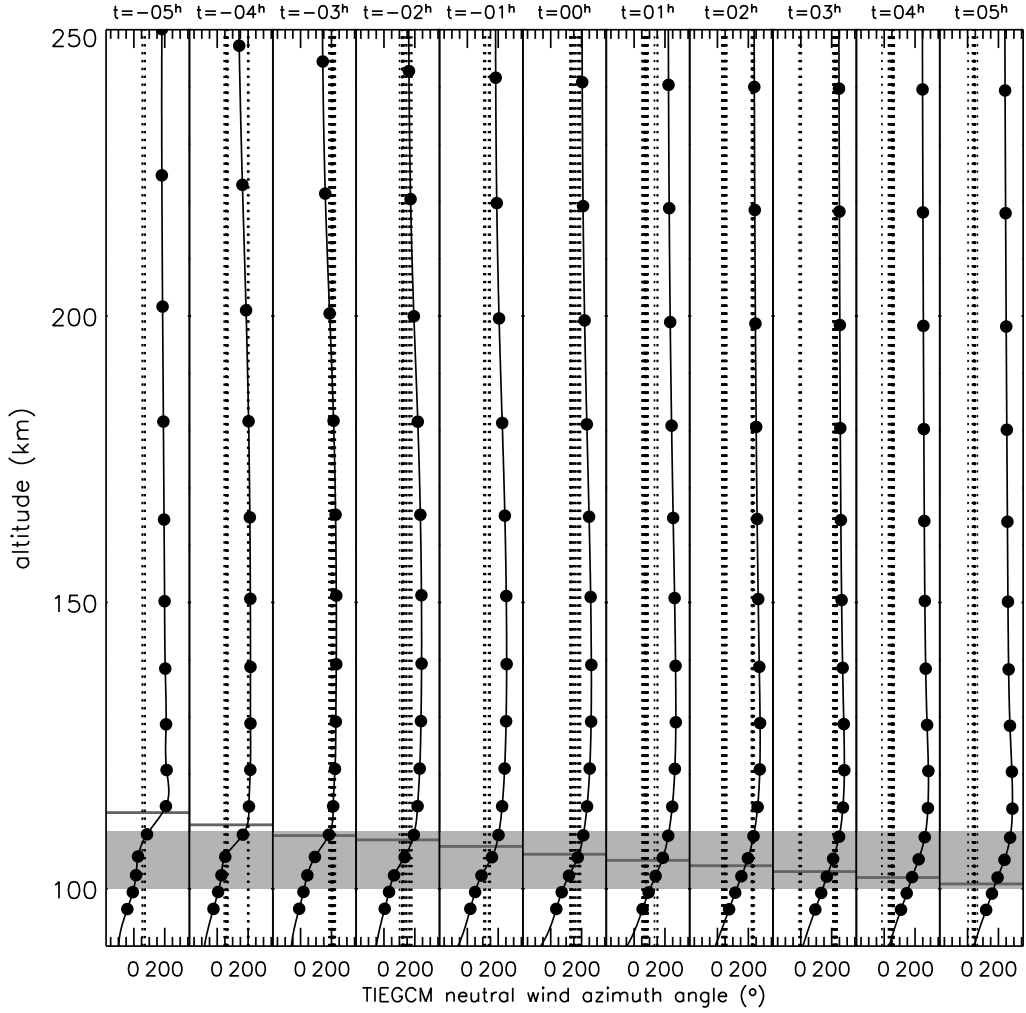}
\caption{Within one-hour windows, the azimuth angle of the neutral wind velocity vector from TIEGCM as a function of altitude (points with cubic spline fits plotted as solid curves).  In each panel, the vertical dotted lines represent the $\vec{k}$ azimuth angles for the $5 \sigma$ detections of small-scale moving structures.  The light-grey shaded regions indicate the typical range in altitudes for $E_s$ layers, 100--110 km.  Each dark-gray horizontal line indicates the altitude where the wind direction changes from eastward at lower altitudes to westward at higher altitudes, i.e., where the wind shear necessary for creating an $E_s$ layer exists.}
\label{windpa}
\end{figure}

The group of small-scale waves seen in Fig.\ \ref{padist} to 
have $\vec{k}$ pointed to the southeast cannot be explained 
with the $E_s$LI.  However, several authors have noted that QP echoes 
are often found with no preferred direction, or with directions 
inconsistent with the direction-dependent $E_s$LI 
\citep[e.g.,][]{fuk98,yam05,lar07}.  \citet{yam05} found a tendency 
for the fronts of QP echoes to be nearly perpendicular to 
the direction of the neutral wind at the height of the $E_s$ layer(s).  To 
explore this possibility, we have plotted the neutral wind velocity azimuth angle profile 
taken from publicly available GPI data-driven runs of the TIEGCM code 
\citep[][currently available at http://www.hao.ucar.edu/modeling/tgcm/]{wan99} 
for the night of 13 August at $32.5^{\circ}$N and $110^{\circ}$W 
at one hour intervals in Fig.\ \ref{windpa}.  
In each panel, the $\vec{k}$ azimuth angles of each $5 \sigma$ 
detection of small-scale waves within the corresponding time bin 
is plotted as a vertical dotted line.\par
To form and maintain an $E_s$ 
layer, a zonal wind shear is required with the westward moving 
wind at higher altitudes than the wind moving eastward.  In the plots 
in Fig.\ \ref{windpa}, this occurs at altitudes where the neutral 
wind transitions from moving southeastward at lower altitudes 
to moving southwestward at higher altitudes.  This usually occurs 
at heights between about 100 and 110 km which is typical for $E_s$ 
layers.  The small-scale waves detected with $\vec{k}$ toward the southeast 
are then consistent with having wavefronts nearly perpendicular to the 
neutral wind at the $E_s$ layer.  We also note that 
the waves with $\vec{k}$ pointing to the southwest which are not coincident with MSTIDs may also be 
consistent with this same scenario, depending on the altitude at which
 the proposed 
$E_s$ layer has formed.\par
\citet{yam05} concluded that these types 
of QP echoes may be formed either by interactions with gravity 
waves or via the Kelvin-Helmholtz instability.  Since these waves were detected during the period of lowest MSTID activity (see Fig.\ 
\ref{armspecpa} and \ref{padist}), $E$--$F$ coupling likely has 
little to do with the generation of these structures.  Gravity waves or Kelvin-Helmholtz instabilities are much more 
likely candidates for their generation mechanism.

\subsection{Background Fluctuations}
\begin{figure}
\noindent\includegraphics[width=6in]{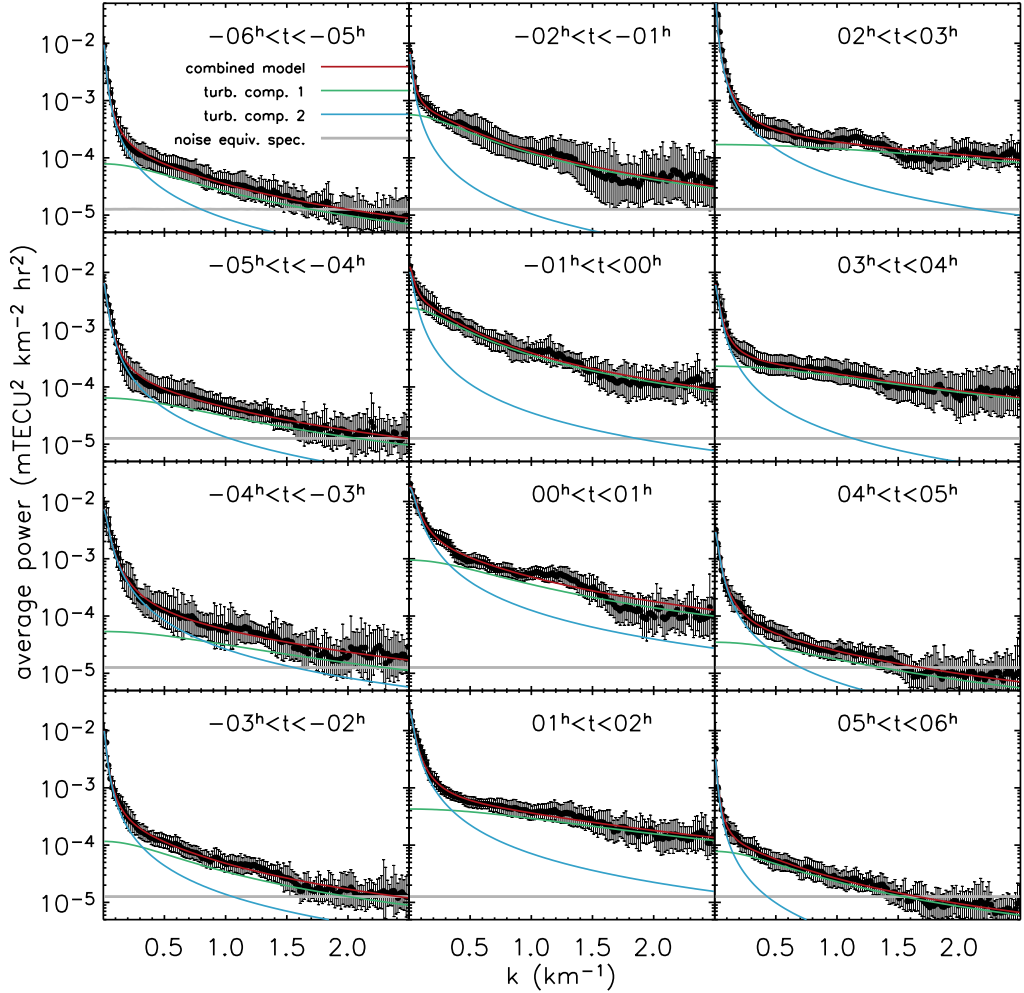}
\caption{Within one-hour bins, the mean estimated background fluctuation spectra (see \S 4.1 and \S 4.4).  The curves represent two-component turbulence fits to the spectra (see \S 4.4; component 1 is the smaller-scale component; component 2 is the larger-scale component).  Noise equivalent spectra are plotted as grey curves.}
\label{armspecbin}
\end{figure}

While our arm-based techniques have found evidence for groups of waves moving 
in preferred directions on all size-scales, these results do not 
exclude the existence of populations of structures moving 
in a wide enough range of directions as to appear isotropic within 
our analysis.  To examine this possibility, we have looked to the background 
spectra constructed using the arm spectra described in \S 4.1.  
These spectra should be reasonably free of the presence of directed waves 
and will provide insights into the quasi-isotropic set of background 
fluctuations.  Mean background spectra computed within one hour bins 
are plotted in Fig.\ \ref{armspecbin}.\par
While some of these spectra 
may be reasonably approximated with single power-laws, many have the 
appearance of ``broken,'' or two-component power-laws.  Upon detailed 
inspection, we found that both the inner and outer most regions of 
the spectra were roughly $\propto k^{-5/3}$, but with different slopes 
and offsets.  This is intriguing since this is the dependence one 
expects for turbulent fluctuations.  In particular, for a thin 
shell, \citet{tat61} predicted from \citet{kol41a,kol41b} that the spectrum of 
refractive index fluctuations gives a spectrum of wavefront phase 
variations $\propto k^{-11/3}$.  Since we have examined the gradient 
of TEC (or, wavefront phase) fluctuations, the fluctuations have 
been effectively multiplied by a factor of $i |\vec{k}|$, and so the 
power spectrum is altered by a factor of $k^2$, or $\propto k^{-5/3}$.  
Thus, the general shapes of the background spectra suggest they may be 
interpreted as the sums of two separate turbulent components.\par
We have consequently fit a two-component turbulent model to each of the 
background spectra plotted in Fig.\ \ref{armspecbin}.  To allow 
each spectrum to flatten above a maximum turbulent scale-length, 
$\lambda_{max}$, we assumed the following form for each component
\begin{equation}
\Psi (k) = \left [ k^2 + \left ( \frac{2 \pi}{\lambda_{max}}\right )^2 \right ]^{- \frac{5}{6}}
\end{equation}
The resulting two-component fits are plotted with the spectra in Fig.\ 
\ref{armspecbin}; the two components are plotted separately as well.  
In general, one can see that each spectrum is well represented by 
a component with a relatively large $\lambda_{max}$ (blue curves) and 
one that flattens at much smaller scales (higher wavenumbers; green 
curves).  To illustrate this quantitatively, we have plotted in Fig.\ 
\ref{armspecfit} the power at $\lambda_{max}$ (upper panel) and 
$\lambda_{max}$ (lower panel) for each of the two components as functions 
of local time. For both components, $\lambda_{max}$ varies somewhat with 
time, but is typically $\sim 300$ km for the larger-scale turbulent 
component and roughly 10 km for the smaller-scale component.  While the 
power at $\lambda_{max}$ seems relatively stable with time for the 
larger component, the small-scale component shows a definite peak at 
a local time of about $-00^{\mbox{\scriptsize h}}30^{\mbox{\scriptsize m}}$.  
This corresponds to the middle of the roughly three-hour lull in 
MSTID activity that started around $-02^{\mbox{\scriptsize h}}$.  Since we also detected small-scale waves consistent with QP-echoes generated by gravity waves during this time, it is possible that this increase in power is the result of the same gravity waves influencing turbulent processes in the lower ionosphere/thermosphere where the role of ion-neutral coupling is relatively strong.

\section{Discussion}
Our exploration of a long, VHF observation of Cyg A with the VLA 
has successfully demonstrated the power of this instrument to 
characterize a variety of transient ionospheric phenomena.  For this 
observation, the typical $1 \sigma$ uncertainty in the $\delta \mbox{TEC}$ 
measurements was $3 \times 10^{-4}$ TECU, yielding more than an order of 
magnitude better sensitivity to TEC fluctuations than can be 
achieved with GPS-based relative TEC measurements 
\citep[see, e.g.,][]{her06}.  Through our new spectral-based analysis, 
we have demonstrated the ability of the VLA to detect and characterize 
individual instances of MSTIDs as well as smaller-scale structures 
likely associated with $E_s$ layers.  We note that since disturbances within the plasmasphere have been observed and in fact discovered with the VLA, it stands to reason that some of the phenomena we have detected may be located within the plasmasphere as well.  However, the phenomena discovered by \citet{jac92b} have azimuth angles clustered around $102^\circ$ (i.e., near magnetic east for the VLA), and few if any of the wave-like structures we have detected meet this criterion.  Therefore, the vast majority (if not all) of the phenomena we have observed were likely within the ionosphere.\par
Among these phenomena are MSTIDs, small-scale wave-like phenomena consistent with $E_s$ layer disturbances, and turbulent fluctuations.  Both the MSTIDs and the small-scale fluctuations were present intermittently throughout the night, and turbulent fluctuations were seen at all times, which is common within VHF observations with the VLA \citep{coh09}.  The MSTIDs appear to change direction after local midnight from being directed generally westward to being directed toward the northeast (see below).  There is also a noticeable change in the turbulent activity near midnight, namely the spectral power of these fluctuations on small ($\sim\!10$ km) scales peaks near midnight and gradually decreases toward dawn.
\subsection{West/northwest-directed MSTIDs}
Before midnight local time, we observed instances of westward/northwestward moving MSTIDs (see Fig.\ \ref{waveprop}) with an additional group moving westward observed briefly near $02^{\mbox{\scriptsize h}}30^{\mbox{\scriptsize m}}$ (see Fig.\ \ref{armspecpa}).  While atypical for what is observed for most of North America \citep[see][]{tsu07}, the directions of these waves are similar to what is typically observed near the west coast in California during summer nighttime \citep[see][]{her06}, especially those directed closer to due west.\par
Small-scale waves were detected coincident with nearly all of these MSTIDs moving toward either the northeast or southwest.  Given the difference in directions, it is unlikely that these are simply the signatures of wavefront distortions within the MSTIDs themselves.  As discussed above, the amplitudes and orientations of these small-scale waves are consistent with them being generated within $E_s$ layers via the $E_s$LI.  Coupling between the $E_s$LI and the Perkins instability in the F region \citep[see][]{cos04a,cos04b} may be what has influenced the MSTIDs observed near dawn to move closer to northward as they were detected concurrently with northeastward-directed small-scale waves.  In contrast, those waves seen near $02^{\mbox{\scriptsize h}}30^{\mbox{\scriptsize m}}$ were moving closer to due west and were accompanied by small-scale waves moving toward the southwest.
\subsection{Northeastward-directed MSTIDs}
Unusual, northeastward-directed MSTIDs were observed near $01^{\mbox{\scriptsize h}}$ to $02^{\mbox{\scriptsize h}}$ and $04^{\mbox{\scriptsize h}}$ to $04^{\mbox{\scriptsize h}}30^{\mbox{\scriptsize m}}$.  As discussed above, \citet{shi08} observed similar phenomena over Japan at midlatitudes during the night when the height of the F region was seen to drop significantly.  Similarly, we have found evidence within data from relatively nearby ionosondes that suggests that the F region may have experienced a similar drop in height during the instances of northeastward-directed MSTIDs observed with the VLA.\par
Small-scale waves were also observed at the same times as these unusual MSTIDs.  They were also mostly directed toward the northeast, suggesting that they may represent small-scale structures within the MSTIDs themselves.  However, we note that (1) Fig.\ \ref{armspeckv} shows that they were moving significantly more slowly than the MSTIDs, (2) the distribution of azimuth angles seen in Fig.\ \ref{padist} to peak near the azimuth angle of the VLA's southwestern arm indicates a bias that only affects groups of waves that span an area smaller than the array, and (3) toward the end of the night, the small-scale waves change direction toward the southwest while the MSTIDs remained northeastward directed.  These factors seem to indicate that the small-scale waves are separate phenomena, similar to the other observed small-scale waves that are consistent with QP echoes generated within $E_s$ layers.  It could be that coupling between these disturbances and the larger waves in the F region plays some role in the unusual direction of the observed MSTIDs.  In any case, a more thorough, statistical examination of these phenomena seems warranted.
\subsection{Southeast/southwest-directed QP Echoes}
\begin{figure}
\noindent\includegraphics[width=6in]{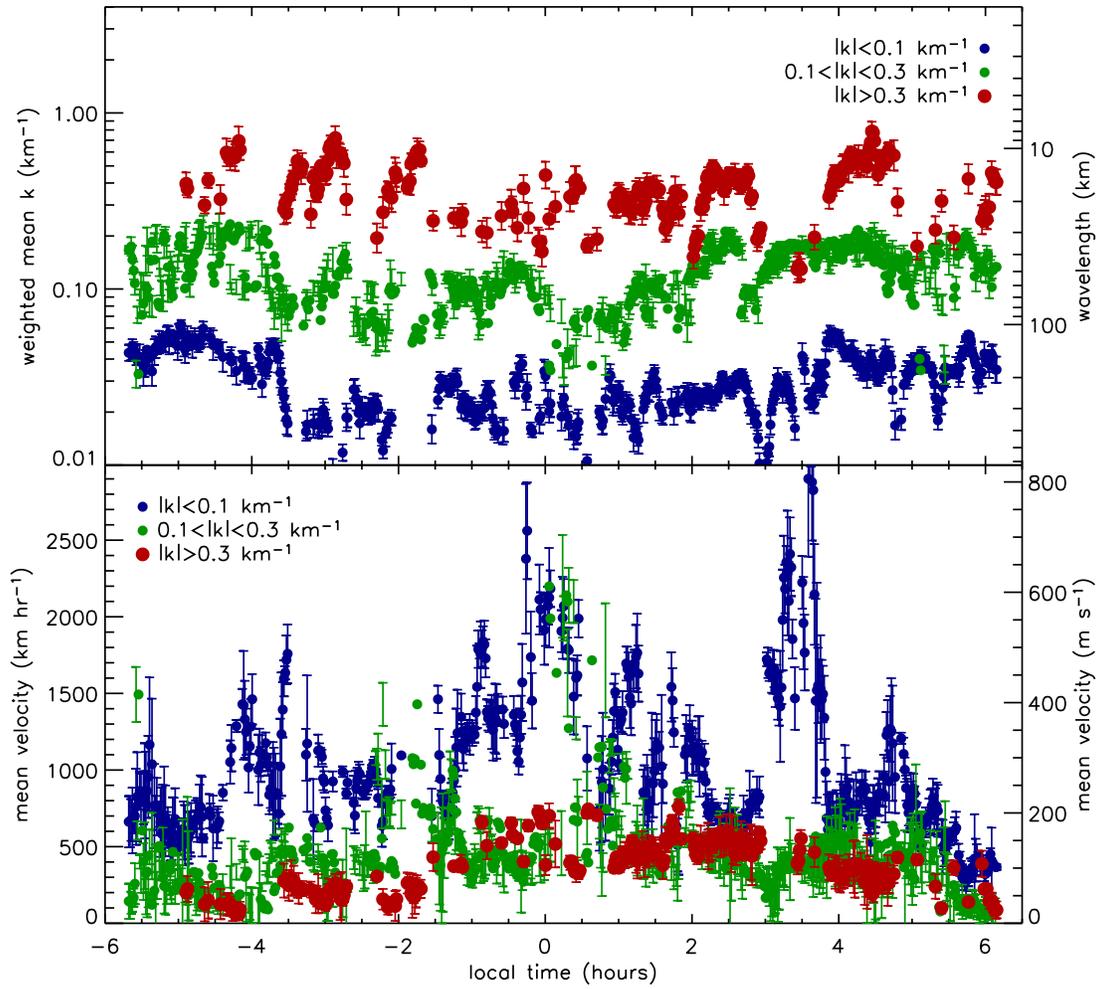}
\caption{The weighted mean wavenumber/wavelength (upper) and velocity (lower) for $5\sigma$ detections (see \S 4.2) for each of the three wavenumber classes.}
\label{armspeckv}
\end{figure}

There are two distinct groups of small-scale waves which were detected with no coincident MSTIDs.  They were predominantly directed toward the southeast (from $-02^{\mbox{\scriptsize h}}$ to $01^{\mbox{\scriptsize h}}$) with a shorter period of activity of waves moving toward the southwest (between $-03^{\mbox{\scriptsize h}}30^{\mbox{\scriptsize m}}$ and $-02^{\mbox{\scriptsize h}}45^{\mbox{\scriptsize m}}$).  The southeastward-directed waves in particular seem to be somewhat different from the other observed small-scale waves.  Given their orientations, they cannot have been generated via the $E_s$LI.  They appear to be moving somewhat faster with 
typical speeds of about 150 m s$^{-1}$ while most of the other small-scale waves have speeds between 30 and 100 m s$^{-1}$ (see Fig.\ \ref{armspeckv}).  They also seem to be largely unaffected by the arm-based bias discussed in \S 4.3, indicating that as a group, the likely span an area larger than that of the array.\par
Despite these differences, the southeastward-directed small-scale waves are generally consistent with QP echoes generated in $E_s$ layers, both in amplitude and wavelength.  They are more specifically consistent with the QP echoes observed by the Sporadic-$E$ Experiments over Kyushu \citep[SEEK][]{fuk98,yam05}, which were heavily influenced by the direction of the neutral wind at the height of the $E_s$ layer(s).  The shift in direction at around $-02^{\mbox{\scriptsize h}}30^{\mbox{\scriptsize m}}$ from southwestward to southeastward may then be an indication of a change in $E_s$ layer height from above the wind sheer altitude to below it.  This may imply the formation/dissipation of $E_s$ layers near this time.\par
\begin{figure}
\noindent\includegraphics[width=6in]{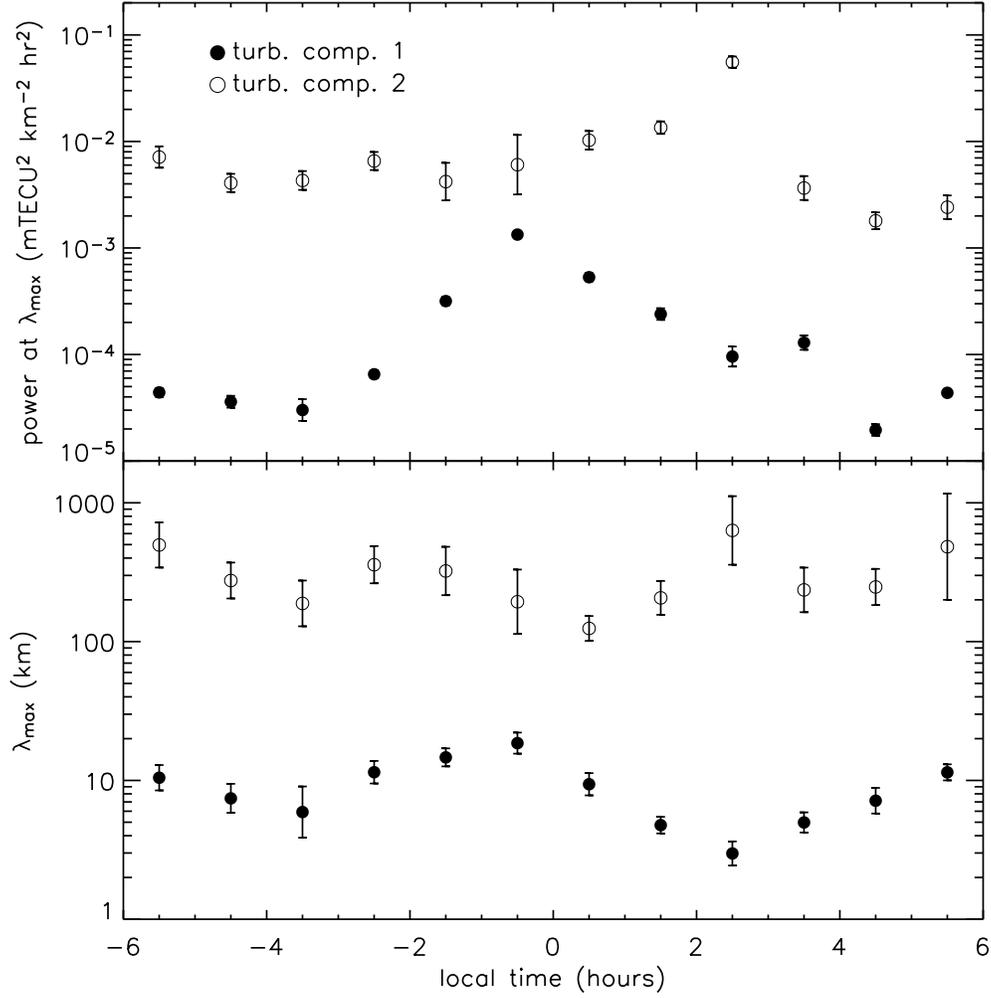}
\caption{For the two-component turbulence fits displayed in Fig.\ \ref{armspecbin}, the power at the maximum wavelength (upper) and the maximum wavelength (lower) for each component as functions of local time [see \S 4.4 and equation (4)].  Here, turbulent component 1 (green curves in Fig.\ \ref{armspecbin}) refers to the smaller-scale turbulent component and turbulent component 2 (blue curves in Fig.\ \ref{armspecbin}) refers to the larger-scale component.}
\label{armspecfit}
\end{figure}

In addition, these waves occurred during the peak in small-scale turbulent activity as evidenced by the results shown in Fig.\ \ref{armspecfit}.  This suggests that the same mechanism that generated these QP echoes also caused a significant turbulent disturbance in the lower ionosphere/thermosphere.  Since the Rocky Mountains lie largely to the north and northwest of the VLA, this mechanism is likely gravity waves associated with wind flow over the Rocky Mountains.

\begin{acknowledgments}
The authors would like to thank the referees for useful comments and suggestions.  Basic research in astronomy at the Naval Research Laboratory is supported 
by 6.1 base funding.  The VLA was operated by the National Radio Astronomy Observatory which is a facility 
of the National Science Foundation operated under cooperative agreement by 
Associated Universities, Inc.  Part of this research was carried out at the Jet Propulsion Laboratory, California Institute of Technology, under a contract with the National Aeronautics and Space Administration.
\end{acknowledgments}

\end{article}

\clearpage

\clearpage

\clearpage

\clearpage

\clearpage

\clearpage

\clearpage

\clearpage

\clearpage

\clearpage

\clearpage

\clearpage

\clearpage

\clearpage

\clearpage

\end{document}